\documentclass[12pt]{article}

\usepackage{cite}

\usepackage[nointlimits,reqno]{amsmath}
\usepackage{amssymb}
\usepackage{simplewick}
\usepackage{slashed}
\numberwithin{equation}{section} 

\usepackage{graphicx}

\usepackage[pdftex,
 pdfproducer=TeXShop,
 pdfcreator=pdflatex]{hyperref}

\usepackage{xspace}
\usepackage{luty}

\begin{document}

\newcommand{\norm}[1]{|\!\gap|#1|\!\gap|}
\newcommand{\aavg}[1]{\avg{\!\avg{#1}\!}}

\begin{titlepage}

\title{Positive Energy Conditions\\
in 4D Conformal Field Theory}

\author{Kara Farnsworth,\ \ Markus A. Luty,\ \ and\ \ Valentina Prilepina}

\address{Physics Department, University of California, Davis\\
Davis, California 95616}

\begin{abstract}

\end{abstract}
We argue that all consistent 4D quantum field theories obey a
spacetime-averaged weak energy inequality
$\avg{T^{00}} \ge -C/L^4$, where $L$ is the size of the smearing region,
and $C$ is a positive constant that depends on the theory. 
If this condition is violated, the theory has states that are
indistinguishable from states of negative total energy by any local measurement,
and we expect instabilities or other inconsistencies.
We apply this condition to 4D conformal field theories,
and find that it places constraints on the OPE coefficients of the theory.
The constraints we find are weaker than the ``conformal collider'' constraints
of Hofman and Maldacena.
We speculate that there may be theories that violate the Hofman-Maldacena
bounds, but satisfy our bounds. 
In 3D CFTs, the only constraint we find is equivalent to the positivity of
2-point function of the energy-momentum tensor, which follows from unitarity.
Our calculations are performed using momentum-space Wightman functions,
which are remarkably simple functions of momenta, and may be of interest
in their own right.
\end{titlepage}

\section{Introduction}
One of the most basic requirements of a consistent quantum field theory
is the existence of a stable ground state of lowest total energy.
Defining the vacuum energy to be zero, we have for all states $\ket\Psi$
\[
\eql{TEC}
\avg{P^0} = \myint d^3 x\, \bra\Psi T^{00}(x) \ket\Psi \ge 0.
\]
This is an example of an energy condition, which requires some
components of the energy-momentum tensor to be non-negative.
We will refer to this as the total energy condition (TEC).
It is expected that any theory that does not satisfy this condition will
have unphysical instabilities.

One may ask whether there are additional energy conditions that hold in
consistent quantum field theories that bound the energy-momentum tensor
locally.
In a classical field theory, it makes sense to impose energy conditions
on the energy-momentum tensor at a spacetime point.
Examples include the weak energy condition (WEC), which states that
$T^{\mu\nu}(x) u_\mu u_\nu \ge 0$ for all timelike vectors $u^\mu$,
and the null energy condition (NEC), which states that
$T^{\mu\nu}(x) n_\mu n_\nu \ge 0$ for all null vectors $n^\mu$.
For example, in a scalar field theory, we have
$T^{00} = \frac 12 \dot{\phi}^2
+ \frac 12 (\vec\nabla\phi)^2 + V(\phi) \ge 0$ provided that the potential
is non-negative.
Therefore, any such theory satisfies the WEC.
In curved spacetime the WEC cannot be a fundamental consistency requirement,
since it is violated by a negative cosmological constant, which
does not give rise to any unphysical instabilities.
However, in this paper we are considering a field theory
in Minkowski space,  with the vacuum energy defined to be zero.
In this context, the WEC is a meaningful condition.
In fact, in a conformally invariant theory where $T^\mu{}_\mu(x) = 0$,
the WEC and NEC are equivalent (see Appendix A).

In a quantum field theory
the expectation value $\avg{T^{00}(x)}$ is not necessarily positive
due to the subtractions that are required to render the energy density finite.
In fact, it can be proved that $\avg{T^{00}(x)}$ is
negative for some states in any quantum field theory \cite{Epstein:1965zza}.
This is because the energy density of the vacuum vanishes,
and therefore must have both positive and negative fluctuations.
This implies the existence of states with
$\avg{T^{00}(x)} < 0$.
In a CFT the energy density can be arbitrarily negative,
since a finite lower bound would violate scale invariance.
We can think of this as a consequence of the uncertainty principle:
the energy is well-defined only to an accuracy given by the time
over which it is measured, so the energy density can have arbitrarily
large fluctuations of either sign.
Similar arguments imply violation of the NEC and all other standard 
pointwise energy  conditions.

We therefore consider energy conditions with some form
of averaging to suppress the fluctuations.%
\footnote{For an interesting attempt to define meaningful local
energy inequalities in quantum field theory, see \Ref{Latorre:1997ea}.}
In this work, we propose a bound on the 
energy density closely related to the positivity of total energy \Eq{TEC}.
It can be defined in terms of the averaged energy-momentum tensor
\beq
T^{\mu\nu}[f] = 
\myint d^4 x\ggap f(x) \ggap T^{\mu\nu}(x),
\eeq
where $f(x)$ is a smooth function with width of order $L$, with
$\int \! d^4 x \ggap f(x) = 1$.
For example, we can use a Gaussian
\beq
f(x) = \scr{N} e^{-\norm{x}^2 / 2 L^2}
\eeq
with $\norm{x}^2 = (x^0)^2 + \vec{x}{}\gap^2$.

Our proposed bound states that 
\beq\eql{SAWEC}
\avg{T^{00}[f]} \ge -\frac{C}{L^4}
\eeq
for some positive constant $C$, for all non-singular states.%
\footnote{A reasonable requirement for a state to be non-singular is
that expectation values of arbitrary
products of local operators are well-defined in such states.}
The constant $C$ depends on the theory and the smearing function; 
for example $C \propto N$ for a large-$N$ theory with the constant of 
proportionality depending on the choice of $f(x)$.
This condition bounds the amount of negative energy seen by
measurements sensitive to the energy averaged over a
spacetime region of size $L$.
We therefore call it the spacetime averaged weak energy condition
(SAWEC).
Note that the SAWEC is saturated by Casimir energy,
so we expect this to be an optimal bound.
Also note that \Eq{SAWEC} is the only possible bound in a scale invariant
theory, as long as the function $f$ depends only on a single scale $L$.

The bound \Eq{SAWEC}
can be viewed as a weaker form of the so-called ``quantum inequalities'' 
(QI) \cite{Ford:1990id,Ford:1994bj,Ford:1996er}.
The particular QI that we will discuss here states that 
for a time sampling function $g(\tau)$ with width $T$ 
(for example, a Gaussian $g(\tau) \propto e^{-\tau^2 / 2T^2}$)
we have for any worldline $x^\mu(\tau)$ parameterized by proper time 
\beq\eql{QI}
\myint d\tau \, g(\tau) \ggap 
\bigl\langle T_{\mu\nu}(x(\tau)) \bigr\rangle
\frac{d x^\mu}{d\tau} \frac{d x^\nu}{d\tau}
\ge -\frac{c}{T^4}.
\eeq
for some positive constant $c$.
This inequality has been proved only for free field theories.
It has been argued that this QI is sufficient to prevent unphysical effects
of macroscopic negative energy (see for example \Refs{Ford:1978qya,Morris:1988tu}).
For a review of other energy conditions, see \Ref{Curiel:2014zba}.

Let us discuss briefly the relation among the three energy conditions discussed
above: the TEC \Eq{TEC}, the SAWEC \Eq{SAWEC}, and the QI \Eq{QI}.
It is not hard to see that
\beq
\text{QI} \Rightarrow \text{SAWEC} \Rightarrow \text{TEC}.
\eeq
Essentially this is because each of the energy conditions in this 
sequence can be viewed as an averaged version of the energy conditions
to the left.
It is easy to see that the QI implies the SAWEC.
We can simply consider a family of worldlines with
$\vec{x} = \text{constant}$, $\tau = t$, and spatially average the quantum 
inequality for these worldlines with a function $h(\vec{x}\gap)$ 
with width $L = T$.
This gives
\beq\eql{AQI}
\myint d^3 x\, h(\vec{x}\gap)
\myint dt \, g(t) \, \avg{T^{00}(\vec{x}, t)}
\ge -\frac{c}{L^4}.
\eeq
This has the form of the SAWEC with $f(x) = h(\vec{x}\gap) g(t)$.
On the other hand, the SAWEC does not immediately imply the QI, 
since there may be theories where some observers see more
negative energy than allowed by the QI, but spatial averaging restores
the bound.
It is also easy to see that the SAWEC implies the TEC.
As $L\to\infty$ the SAWEC is averaging over all of space.
The averaging over time does not matter in this limit because $\avg{P^0}$
is independent of time.
We therefore have as $L \to \infty$
\beq
\myint d^4 x\, f(x) \avg{T^{\mu\nu}(x)}
\to \avg{P^0} \ge 0.
\eeq
The TEC does not imply the SAWEC, at least not in any elementary way.
The reason is that it is possible to have a region with arbitrarily large
negative energy violating the SAWEC, with compensating positive energy
outside the region so that the TEC is satisfied.
In other words, theories that violate the SAWEC but satisfy the TEC
are those where positive and negative energy can be arbitrarily
separated.

There are strong reasons for believing that the SAWEC 
is a fundamental requirement in any consistent quantum field theory.
In any quantum field theory, for sufficiently large $L$ we would expect that
any violation of the SAWEC would show up in the effective field
theory describing the long-distance limit of the field theory.
In a scale invariant theory, the only possible bound for the averaged
energy density is the SAWEC, so if it is violated 
one can put an arbitrarily large amount of negative energy in any size
region.
By taking the size $L$ arbitrarily large, one can construct a state
that is indistinguishable from a state of negative total energy by any 
local measurement.
It seems very likely that such theories in fact have states of negative 
total energy, or some other inconsistency.

With this background and motivation,
we study positivity of energy in 4D conformal field theories (CFTs).
We will consider states defined by acting on the vacuum with a local operator.
Specifically, we focus on states of the form
\beq\eql{wavepacket}
\ket\Psi = \myint d^4 x\ggap e^{i p \cdot x}  e^{-\norm{x}^2/R^2}
\ggap \ep_{\mu\nu} T^{\mu\nu}(x) \ket 0\,,
\eeq
where
$p$ is a 4-momentum, $\ep_{\mu\nu}$ is a polarization tensor,
and $R \gg 1/p$ is a long-distance cutoff.
The energy-momentum tensor $T^{\mu\nu}$ is present in any local conformal field theory,
so this state exists in any such theory.
We take $R \gg 1/p$ so that the total momentum $P^\mu$
is approximately well-defined in this state. 
We are working in Heisenberg picture in Minkowski space, 
so $\ket\Psi$ describes the system for all times.
On the time slice $x^0 = 0$ the state can be thought of as a wavepacket 
centered at $\vec{x} = 0$ with radius $R$ and average 
4-momentum $p^\mu$.
We then consider the energy-momentum tensor evaluated in the state \Eq{wavepacket}:
\beq
\avg{T^{\mu\nu}(x)} = \frac{\displaystyle \bra\Psi T^{\mu\nu}(x) \ket\Psi}
{\displaystyle \braket\Psi\Psi}.
\eeq
This is a local quantity, but as $R \to \infty$
$\ket\Psi$ approaches an eigenstate of $P_\mu$ and
$\avg{T^{\mu\nu}(x)}$ becomes independent of $x$.
This means that we are effectively evaluating a smeared quantity.
More precisely, for a smearing function with $L$ held fixed as
$R \to \infty$ we have
$\avg{T^{\mu\nu}[f]} \to \avg{T^{\mu\nu}(0)}$.
Therefore enforcing the SAWEC in this context is equivalent to requiring 
the local condition $\avg{T^{00}(0)} \geq 0 $.

The expectation value of the energy-momentum tensor in the state \Eq{wavepacket}
is is given by an integral over a 3-point function of the energy-momentum tensor
in a conformal field theory.
This 3-point function is determined by conformal invariance up to 3 OPE coefficients
\cite{Osborn:1993cr}.
We perform the calculation using the momentum-space Wightman functions.
We show that these are surprisingly simple functions of momenta,
suggesting that they may be useful in other applications.
By requiring the SAWEC, we find 2 constraints among these 3 parameters
(specified in \Eq{ourconstraints} below).
One of these constraints is equivalent to the positivity of the $c$ anomaly coefficient,
which follows from unitarity.

If these constraints are violated,
we have constructed a state that has arbitrarily
negative constant energy density in a region of any size.
On the other hand, the total energy is always positive in the states we
construct.
(In fact, it is built into our calculation that $P^0$ has only
positive eigenvalues in the state \Eq{wavepacket}.)
What is therefore happening is that positive and negative energy densities can be 
arbitrarily separated,  with negative energy density in the interior of the wavepacket, 
but enough positive energy density outside the wavepacket to ensure that the total 
energy is positive.
By taking the size of the smearing region arbitrarily large, we can construct a state that
is indistinguishable from a state of constant negative energy density by any
local measurement.
It is hard to believe that this does not signal an instability, especially since
operator insertions are giving rise to negative energy density in the region where
the operators are inserted.
It seems very likely that such theories have runaway
instabilities where positive energy is radiated to infinity,
while the local energy density decreases without bound.
It would be very interesting to explicitly exhibit such
an instability.
Interestingly, states with large positive energy density in a region with negative
energy radiated to infinity can be rigorously ruled out using conformal invariance and 
positivity of total energy \cite{Blanco:2013lea}.

Our bounds are closely related to (and were inspired by)
the work of Hofman and Maldacena (HM) in \Ref{Hofman:2008ar}.
They also considered states of the form \Eq{wavepacket}, but
calculated a different observable:
the energy flux $\scr{E}$ measured by a calorimeter cell at infinity
(see Fig.~1).
This is also determined by the 3-point function of the energy-momentum
tensor, and HM showed that the condition 
$\scr{E} \ge 0$ also gives nontrivial constraints on the $TTT$
OPE coefficients  of the theory.
The constraints obtained by HM in this way are stronger than our constraints.
For example, in $\scr{N} = 1$ supersymmetric theories there are only 2
independent $TTT$ OPE coefficients, and we find
\[
-4 \le \frac{a}{c} \le 6.
\]
For comparison, the constraints of Hofman and Maldacena in this case are
$\frac 12 \le a/c \le \frac 32$.
Additional constraints arising from positivity of higher point
energy correlation functions are discussed in \Ref{Zhiboedov:2013opa}.

We also considered the bounds in 3D CFTs, where we find no bounds at
all, while the HM constraints are still nontrivial in this case.
This raises the question of whether there are theories that violate
the HM constraints, but satisfy our bounds.

The condition that the energy flux measured by a distant observer
is positive certainly sounds physically reasonable, but it not 
clear that it is really necessary for consistency of the theory. 
The total energy of the states \Eq{wavepacket} is always positive,
so the novel feature of the HM constraints is that a local
perturbation creates a state that takes energy from some asymptotic
region.
This is certainly peculiar, but it violates neither causality nor
positivity of energy.
Negative energy flux can be measured by physical calorimeters by
a process of stimulated de-excitation.
It is not obvious to us that there is any inconsistency with having
negative expected values for the energy flux.

One case in which the HM bounds appear well-motivated is a
conformal field theory with a relevant perturbation that causes it to 
flow to a theory with a mass gap $m$ in the IR.
Such theories are similar to QCD in that they are conformal in the UV,
but have a well-defined $S$-matrix with asymptotic states of free particles
(``hadrons'').
It is reasonable to expect that for events with total energy
$E \gg m$ the energy flux of the
(presumably many) massive particles at infinity approximates that of the
unperturbed CFT, and therefore the expected value of the energy
flux of the unperturbed CFT should be positive in every direction.
This assumption is known to work well in QCD, where the distribution
of the energy flux predicted by perturbative QCD is in good agreement
with the distributions of hadrons measured at $e^+ e^-$ colliders 
\cite{Abreu:1990us,Akrawy:1990hy,Abe:1994wv}.
In order for a gapped conformal theory to violate the HM constraints,
there would have to be ``hadronization'' effects that do not vanish
as $E \to \infty$.
It is not inconceivable that such theories may exist.
For example, there are large hadronization effects in QCD at large $N_\text{c}$,
where a state of the kind we are considering produces a single highly
excited meson at rest, which we can think of as a quark-antiquark pair
connected by a QCD string.
String breaking is suppressed by $1/N_\text{c}$, so the distribution of
energy at infinity is strongly affected by the details of hadronization. 
With this example in mind, we should be cautious in assuming that energy flux 
in the CFT is positive even in theories that can be gapped.
We should also remember that 
these considerations tell us nothing about theories that
do not have a deformation to a gapped theory, for example CFTs with
no relevant operators.
It may be that the theories that violate the HM constraints are 
restricted to this class of theories.

HM also showed that in a conformal theory the energy flux at
infinity is given by
\[
\scr{E} = \myint dx^-\, \avg{T_{--}(x)}.
\]
where the integral is taken over a future null line at the angular 
position of the calorimeter.
Therefore, the condition that $\scr{E} \ge 0$ is precisely the averaged null energy condition
(ANEC). This is suggestive, since the classical ANEC is sufficient to guarantee the absence of
several kinds of pathological behavior in general relativity \cite{Penrose:1993ud,Gao:2000ga}.
However, the ANEC has been proven only for free field theory 
\cite{Klinkhammer:1991ki,Wald:1991xn}
and for holographic theories \cite{Kelly:2014mra}.
An interesting general argument for the ANEC is given in \Ref{Hofman:2009ug}, 
but it relies on a continuation of lightcone coordinates to Euclidean
space that is not clearly justified, 
and also requires some assumptions about non-local operators.%
\footnote{The ANEC can be derived from the QI by taking null limits of
timelike trajectories.
It is not clear whether one can derive the SAWEC from the ANEC because the latter
involves an integral over an infinite trajectory.}

\begin{figure}[htbp!]
\begin{center}
\includegraphics[scale=0.8]{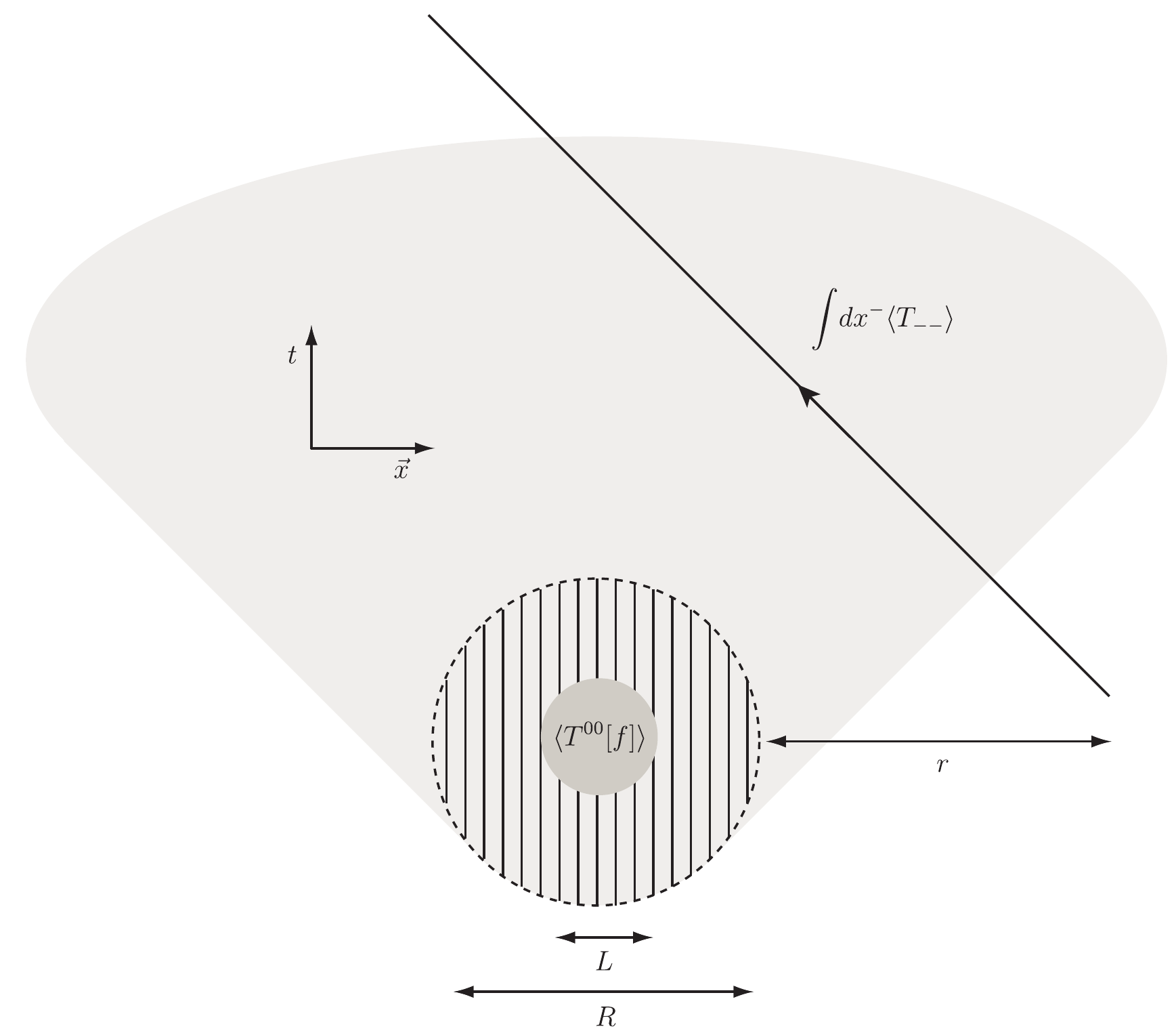} 
\begin{minipage}{5.5in}
\caption{Spacetime diagram illustrating the energy conditions discussed in the text.
In all cases, the wavepacket state is defined by acting with the energy-momentum tensor 
smeared in spacetime over a distance $R \gg 1/p$, illustrated by the hatched region.
This disturbance then spreads out as shown in the shaded region.
The SAWEC is defined by evaluating the energy-momentum tensor smeared over a distance
$L \ll R$ inside the wavepacket.
The ANEC is defined by averaging the energy-momentum tensor over a null line
at a distance $r \gg R$ from the wavepacket.}
\label{fig:HMvsus}
\end{minipage}
\end{center}
\end{figure}

Let us compare our bounds with those of HM at a conceptual level.
If either our conditions or those of HM are violated,
there are processes in which positive and negative energy can be separated,
raising the question of runaway instabilities.
The crucial difference in our view is that violating our conditions implies that
negative energy is created directly by the action of local operators at 
the location of the operator insertion,
while the violation of the HM conditions implies that local operators
can set up negative energy flux at infinity.
In neither case is it demonstrated that violation of the condition
leads to instabilities.  
However, it is harder to imagine how instabilities can be avoided if our 
conditions are violated, since local perturbations can directly lower the 
local energy density at the location of the perturbations.
We therefore believe that it is quite possible 
that there are physically sensible theories that 
violate the HM conditions, but satisfy our conditions.
It would be very interesting to shed further light 
on the question of instabilities in 4D CFTs violating
various energy conditions. 

This paper is organized as follows. 
In \S 2 we give the general form of the expectation value of the energy-momentum tensor 
in a state of the form \Eq{wavepacket} and derive our main result, 
constraints on the OPE coefficients that come from requiring positivity of 
the energy density for both 4D and 3D CFTs. 
In \S 3 we provide the essential ideas and results on the calculations of the Wightman 3-point functions that lead up to our constraints. 
We conclude in \S 4, where we discuss some open questions and directions for future work. 
In Appendix A we present a concise proof of the equivalence of the WEC and NEC for CFTs. 
In Appendix B we provide complete expressions for the momentum-space Wightman 
3-point functions used in this work.

\section{Negative Energy from 3-Point Functions}
We are considering wavepacket states of the form \Eq{wavepacket}.
We compute the expectation value of $T^{\mu\nu}(x)$ for points $x$
well inside the wavepacket ($\norm{x} \ll R$).
This is determined by the Wightman 3-point function
of the energy-momentum tensor,
$\avg{T^{\mu\nu}} \propto \ep^\ast_{\rho\si}\ep_{\tau\om}\avg{T^{\rho\si}T^{\mu\nu}T^{\tau\om}}$.
This 3-point function has 3 independent conformally invariant tensor structures,
and is therefore determined by 3 independent constants \cite{Osborn:1993cr}.
Two of these may be taken to be the coefficients of the 4D Weyl anomaly, $c$ and $a$.
A convenient parameterization of all 3 constants uses the fact that there
are 3 different free CFTs in 4D: real scalar, Dirac fermion, and free vector.
These span the possible tensor structures of the 3-point function, so we can
parameterize the 3-point function in terms of
the coefficients of these 3 tensor structures, which we call
$n_s$, $n_f$, and $n_v$ respectively.
In a free CFT, these are the number of scalars, fermions, and vectors, but they are arbitrary constants in a general CFT.

For $R \to \infty$, the expectation value of the energy-momentum tensor
has the general form
\[
\eql{rhoform}
\avg{T^{\mu\nu}}
&= \frac{\bra\Psi T^{\mu\nu}(0) \ket\Psi}{\braket\Psi\Psi}
\nonumber\\
&= \frac{\sqrt{p^2}}{R^3} \Bigl[
A \frac{(\ep^* \ep)^{(\mu\nu)}}{ \tr(\ep^* \ep)} + B \frac{p^\mu p^\nu}{p^2}
+ C  \eta^{\mu\nu} \Bigr],
\]
where 
\[
(\ep^* \ep)^{(\mu\nu)} = \sfrac 12 \eta_{\rho\si} \ep^{*\mu\rho} \ep^{\si\nu}
+ ( \mu \leftrightarrow \nu),
\qquad
\tr(\ep^* \ep) = \eta_{\mu\nu} \eta_{\rho\si}  \ep^{*\mu\rho} \ep^{\si\nu}.
\]
\Eq{rhoform} is the most general symmetric tensor structure that is quadratic in the polarization tensors, depends on the metric and a single momentum $p^\mu$, and obeys the generality-preserving condition $p^\mu \epsilon_{\mu\nu} = 0$, which follows from the conservation of the energy-momentum tensor.

Tracelessness of the energy-momentum tensor gives%
\footnote{The conformal anomaly gives a local contribution to the 3-point function
that does not contribute to the Wightman function.}
\[
C = - \frac 14 (A + B).
\]
A lengthy computation (described in detail below) gives
\[
A &= \al (-n_s - 3 n_f + 12 n_v),
\\
B &= \al (2 n_s + 13 n_f + 32 n_v),
\]
where
\[\eql{alpha}
\al = \frac{\sqrt{2}}{7 \pi^{3/2}} \frac{1}{(n_s + 6 n_f + 12 n_v)}.
\]

We use these results to compute the energy density in our state.
We find it convenient to boost to a frame where 
\beq
p^\mu = (E, 0, 0, 0).
\eeq
In this rest frame, $u^\mu$ takes the general form
\beq
u^\mu = \bigl(\cosh(y), \sinh(y), 0, 0 \bigr),
\eeq
so the energy density is
\beq
\avg\rho = \avg{T^{\mu\nu}} u_\mu u_\nu.
\eeq
Because $T^{\mu\nu}$ is conserved, symmetric and traceless,
only the spatial components of $\ep_{\mu\nu}$ contribute to the
matrix element, and we can choose without loss of generality
\beq
\ep_{\mu\nu} = \begin{pmatrix}
0 & 0 & 0 & 0 \\
0 & \ep_{11} & \ep_{12} & \ep_{13} \\
0 & \ep_{12} & \ep_{22} & \ep_{23} \\
0 & \ep_{13} & \ep_{23} & \ep_{33}
\end{pmatrix},
\qquad
\ep_{11} + \ep_{22} + \ep_{33} = 0.
\eeq
We then have
\beq
\eql{rhoexpr}
\avg\rho = \frac{\sqrt{p^2}}{R^3} 
\Bigl[ 
-A r \sinh^2(y)
+ B \cosh^2(y) 
+ C
\Bigr],
\eeq
where
\beq
r = -\dfrac{(\ep^\ast \ep)_{11}}{\tr(\ep^\ast\ep)}.
\eeq
Taking into account the tracelessness of the polarization tensor,
the allowed range of the parameter $r$ is
\beq
0 \le r \le \sfrac 23.
\eeq
The limits can be saturated with a diagonal polarization tensor;
the upper limit corresponds to $\ep_{11} = 0$ 
while the lower limit corresponds to 
$\ep_{11} = 1$, $\ep_{22} = \ep_{33} = -\frac 12$.
We see that positivity of the energy density depends only on the two parameters
$y$ and $r$.

We now find the constraints that result from requiring $\avg{\rho} \ge 0$
for all values of $y$ and $r$.
In the limit $y \to \infty$ we have
\[
-A r + B \ge 0.
\]
For $r = 0$, this gives
\[
B \ge 0 
\]
while for $r = \sfrac 23$ we obtain
\[
B \ge \sfrac 23 A.
\]
It is not hard to see that if these constraints are satisfied,
we have $\avg{\rho} \ge 0$ for all choices of $y$ and $r$.
In this way, we obtain our main result
\[
\eql{ourconstraints}
\begin{split}
 2n_s + 13 n_f +32 n_v &\ge 0,
\\
8n_s + 45 n_f+72 n_v  &\ge 0.
\end{split}
\]
One linear combination of these constraints implies the positivity
of $c$:
\[
c \propto n_s+ 6 n_f  + 12 n_v \ge 0.
\]
This was already known from the fact that the 2-point function of
$T^{\mu\nu}$ is proportional to $c$, 
which must be positive in a unitary theory.
We therefore have 1 new constraint.
For comparison, Hofman and Maldacena obtained 3 constraints,
which can be written simply and suggestively as
\[
n_s, n_f, n_v \ge 0. 
\]
One of these is also equivalent to $c \ge 0$, so HM obtain 2 new constraints that
are manifestly stronger than our constraints \Eq{ourconstraints}.

Note that our constraints come entirely from 
the limit $y \to \infty$, where $u^\mu$ approaches a null vector.
This means that our constraints could equivalently be derived from requiring
that the expectation value of the energy-momentum tensor obeys the null
energy condition (NEC):
\beq
\avg{T^{\mu\nu}} n_\mu n_\nu \ge 0
\eeq
for all null vectors $n^\mu$.
In fact, violation of the NEC is equivalent to the existence of a reference
frame with $T^{00} < 0$, as shown in Appendix A.

Although the energy density inside the smearing region is negative 
if our bounds are violated,
we know that the total energy of the state is positive.
This can be understood easily from the Wightman functions in momentum
space, which have the form (see \S 3 and the Appendix B for more details)
\[
\bra{0} \tilde{\scr{O}}(p_f) \tilde{\scr{O}}(-q) \tilde{\scr{O}}(-p_i)
\ket{0} \propto \th(p_i^2)\th(p_i^0) \th(p_f^2) \th(p_f^0).
\]
The step functions imply that the only intermediate states that
contribute to the 3-point function have timelike momentum and positive energy.
(Note that these are Lorentz-invariant conditions.)
Therefore, a state of the form we are considering is a linear combination
of positive-energy states, and must have positive energy.

We have also performed the same calculation in 3D CFTs.
In this case, there are only 2 independent tensor structures in the
3-point function of the energy-momentum tensor,
which can be parameterized by 2 free CFTs: a free real scalar
and a free Dirac fermion.
The expectation value of the energy-momentum tensor is again given by \Eq{rhoform} with the obvious
change $R^3 \to R^2$, and with
\[
C = -\frac 13 (A + B).
\]
The computation of the 3-point function gives
\[
A &= -4 \beta (n_s + n_f ),
\\
B &= \beta (7 n_s + 16 n_f ),
\]
where
\[\eql{beta}
\beta = \frac{1}{6\sqrt{2} \pi} \frac{1}{(n_s + 2 n_f)}.
\]
The energy density is given by \Eq{rhoexpr} with $r$ given by
the obvious restriction to 3D.
In this case we have $r = \sfrac 12$, and we obtain only one constraint
\[
B  \geq \frac 12 A.
\]
This translates to 
\[
n_s + 2 n_f  \geq 0
\]
which is just the positivity of the coefficient of the two point function in 3D. 
We therefore obtain no new constraints in the 3D case.

\section{Wightman Functions in Momentum Space}
In this section we describe the computation of the Wightman 3-point functions
used to derive the results of the previous section.
We have attempted to present the main ideas and results, and some
of the more gory details are relegated to Appendix B. Although we take the limit $R \to \infty$ to find our final results, we want to emphasize that the 3-point functions derived in Appendix B are not approximations and are true in general.
We are interested in the 3-point function of the energy-momentum tensor in a
general 4D CFT.
The general form of the 3-point function in position space was first worked
out by Osborn and Petkou \cite{Osborn:1993cr}.
This has 3 tensor structures, corresponding to the 3 different free CFTs in 4D:
a free real scalar, a free Dirac fermion, and a free vector.
We can therefore write
\beq
\avg{TTT} = n_s \avg{TTT}_s + n_f \avg{TTT}_f + n_v \avg{TTT}_v,
\eeq
where $\avg{TTT}_s, \avg{TTT}_f, \avg{TTT}_v$ are the Wightman 3-point functions for
the theory of a single free scalar, fermion, and vector, respectively.%

The Wightman 3-point function is defined by a  $i\ep$ prescription in 
position space \cite{Luscher:1974ez}.
We choose to compute the Wightman functions directly in momentum space
for the 3 different free field theories.
From these it is straightforward to evaluate the expectation value of
the energy-momentum tensor in the state \Eq{wavepacket}.
We have
\[
\eql{wavepacketint}
\bra\Psi T^{\mu\nu}(0) \ket\Psi
= \myint \frac{d^4 k}{(2\pi)^4} \myint \frac{d^4 \ell}{(2\pi)^4} \,
\tilde{f}^*(\ell) \tilde{f}(k) 
\aavg{\tilde{\scr{O}}(\ell) \tilde{T}^{\mu\nu}(k - \ell) 
\tilde{\scr{O}}(-k)},
\]
where $\scr{O} = \ep_{\rho\si} T^{\rho\si}$ and $\aavg{\cdots}$ is the 
Wightman function in momentum space
with the energy-momentum conserving delta function factored out:
\beq
\begin{split}
\myint d^4 x_1 \, e^{i p_1 \cdot x_1}  
\cdots \myint d^4 x_n \, & e^{i p_n \cdot x_n} \,
\bra{0} \scr{O}_1(x_1) \cdots \scr{O}_n(x_n) \ket{0}
\\
&= (2\pi)^4 \de^4(p_1 + \cdots + p_n) 
\aavg{ \tilde{\scr{O}}_1(p_1) \cdots \tilde{\scr{O}}_n(p_n)}.
\end{split}
\eeq
The function $\tilde{f}(k)$ is the Fourier transform of the wavepacket profile
defined in \Eq{wavepacket}:
\[
\tilde{f}(k) &= \myint d^4 x\, e^{i k \cdot x}
\, e^{-i p \cdot x} e^{-\norm{x}^2 / 2 R^2}
\nonumber\\
&= \tilde{\scr{N}} e^{-\norm{k - p}^2 R^2 /2}.
\]
The normalization factor $\tilde{\scr{N}}$ drops out of our final results.
The Wightman 3-point function in \Eq{wavepacketint} is a simple function 
of the momenta, and we are able to perform the integrals in \Eq{wavepacketint}
explicitly.

\subsection{Free Scalars}
We first describe the computation of the Wightman 3-point function for
the case of scalars.
We work in $d$ spacetime dimensions and will specialize to $d = 3,4$ later.
The (improved) energy-momentum tensor for a scalar $\phi$ is given by
\[
\eql{Tscalar}
T_{\mu\nu} &= \partial_\mu \phi \partial_\nu \phi 
- \sfrac{1}{2} \eta_{\mu\nu} (\partial\phi)^2 
- \sfrac 12 \xi (\partial_\mu \partial_\nu - \eta_{\mu\nu}\Box)\phi^2,
\]
where
\[
\xi = \frac{d - 2}{2(d-1)}.
\]
Computing the 3-point function therefore requires us to calculate the 3-point
functions of the operators 
\beq
\eql{AB}
A_{\mu\nu} = \d_\mu \phi \d_\nu \phi
\quad \textnormal{ and } \quad
B_{\mu\nu} = \d_\mu \d_\nu \phi^2.
\eeq
These can be determined by a simple set of diagrammatic rules.
We define the contraction of scalar fields by
\[
\bcontraction{}{\phi}{(x) }{\phi}
\phi(x) \phi(y) = \bra{0} \phi(x) \phi(y) \ket{0}.
\]
The Wightman functions are then given by a sum of contractions, just as in
the familiar time-ordered perturbation theory.
Because of translation invariance, these rules are simpler in momentum space,
where we have the Wightman propagator
\[
W(p) = \myint d^d x\, e^{i p \cdot x} \, \bcontraction{}{\phi}{(x) }{\phi}
\phi(x) \phi(0)
= 2\pi \de(p^2) \th(p^0)
\]
That is, the propagator is a Lorentz-invariant delta function that puts the
momentum of an internal line on the mass shell.
For example, the 3-point function of the operator $\scr{O} = \frac 12 \phi^2$ is 
\[
\eql{W3O}
\aavg{\tilde{\scr{O}}(p_1) \tilde{\scr{O}}(p_2) \tilde{\scr{O}}(p_3)}
= \myint \frac{d^d k}{(2\pi)^d}\, W(k)
\, W(p_i - k) W(p_f - k),
\]
where $p_f = p_1$, $p_i = -p_3$.
This is an integral over on-shell momenta that is very similar to a phase
space integral.
The momentum of each internal line is put on shell, so the diagram
only contains contributions from real intermediate states.
There is a maximum value of energy and momentum for internal lines, 
so there is no UV divergence.
A straightforward calculation gives
\[
\aavg{\tilde{\scr{O}}(p_1) \tilde{\scr{O}}(p_2) \tilde{\scr{O}}(p_3)}
= \begin{cases}
\displaystyle
\frac{\th(p_i^0) \th(p_i^2) \th(p_f^0) \th(p_f^2)\th(-q^2)}
{2(-p_i^2 p_f^2 q^2)^{1/2}}
& 
d = 3,
\\[15pt]
\displaystyle
\frac{\th(p_i^0) \th(p_i^2) \th(p_f^0) \th(p_f^2)\th(-q^2)}
{8\bigl[ (p_i \cdot p_f)^2 - p_i^2 p_f^2 \bigr]^{1/2}}
& d = 4,
\end{cases}
\]
where $q = p_f - p_i$,
and $\th$ is a step function.
The first four step function factors in the numerators enforce the condition
that the intermediate states have timelike momentum 
with positive energy.
The factor of $\th(-q^2)$ does not follow from requiring physical
intermediate states.
It can be understood as a consequence of the kinematics of the triangle diagram:
$q^2 < 0$ is required so that all three propagators in the triangle diagram are on shell.
The fact that the 3-point function of a dimension-2 scalar operator is fixed
by conformal invariance means that this factor is present in a general CFT.
It would be nice to have a more general understanding of this structure.

To compute the energy-momentum tensor 3-point function for the free scalar,
we need to determine the 3-point functions $\aavg{AAA}$, $\aavg{BAA}$, $\aavg{ABA}$, $\aavg{ABB}$, $\aavg{BAB}$, $\aavg{BBB}$, where $A$ and $B$ are defined in \Eq{AB}, and take the appropriate linear combinations.
Note that all of these integrals are UV finite, so there is no conformal
anomaly and the energy-momentum tensor is traceless.
Each one of these is an integral similar to \Eq{W3O}, for example
\[
\eql{resS}
&\aavg{\tilde{A}_{\mu\nu}(p_1) \tilde{A}_{\rho\si}(p_2) \tilde{A}_{\tau\om}(p_3)}
\nonumber\\
&\qquad\qquad{}
= \myint \frac{d^d k}{(2\pi)^d} \,
k_{(\mu} (p_f - k)_{\nu)}
(p_f - k)_{(\rho} (p_i - k)_{\si)}
(p_i - k)_{(\tau} k_{\om)}
\nonumber\\
&\qquad\qquad\qquad\qquad\quad{}
\times
W(k) W(p_f - k) W(p_i - k).
\]
Complete results for the momentum-space Wightman functions of the operators
$A_{\mu\nu}$ and $B_{\mu\nu}$ are presented in Appendix B. 


Computing these 3-point functions requires us to evaluate integrals of the form
\[
\eql{Imu}
I^{(d)}_{\mu_1 \dots \mu_n}(p_i, p_f) = \myint \frac{d^d k}{(2\pi)^d} 
\, k_{\mu_1} \cdots k_{\mu_n} W(k) W(p_i-k) W(p_f-k)
\]
with values of $n$ up to 6.
It is useful to define a generating function for these diagrams:
\[
\eql{Gen}
I^{(d)}(p_i,p_f; y) = \myint \frac{d^d k}{(2\pi)^d}\, e^{k \cdot y}\ W(k) W(p_i-k) W(p_f-k).
\]
This can be evaluated using standard integral identities and differentiated appropriately with respect to $y$ to obtain the necessary expressions for \Eq{Imu}.
For $d = 3$ and 4 we obtain respectively
\[
I^{(3)}(p_i,p_f; y) &= \th(p_i^0) \th(p_i^2) \th(p_f^0) \th(p_f^2)\th(-q^2)
\times J^{(3)}
e^{K \cdot y} \cosh\bigl(\sqrt{X} \bigr)
\\
I^{(4)}(p_i,p_f; y) &= \th(p_i^0) \th(p_i^2) \th(p_f^0) \th(p_f^2)\th(-q^2)
\times J^{(4)}
e^{K \cdot y} I_0\bigl(\sqrt{X} \bigr)
\]
where, 
\[
J^{(3)} &= \frac{1}{2(-p_i^2p_f^2 q^2)^{1/2}},
\\
J^{(4)} &= \frac{1}{8D^{1/2}},
\\
K^\mu &= \frac{1}{2D} \left[
p_f^2 (q \cdot p_i ) p_i^\mu
- p_i^2 (q \cdot p_f ) p_f^\mu \right],
\\
D &= (p_i\cdot p_f)^2-p_i^2p_f^2,
\\
X &= \frac{p_i^2 p_f^2 q^2}{4D} Y^\mu Y_\mu,
\\
Y^\mu &= y^\mu + \frac{1}{D^{1/2}} \bigl[
(p_f \cdot y) p_i^\mu 
- (p_i \cdot y) p_f^\mu
\bigr],
\]
and $I_0$ is a modified Bessel function of the first kind.
Note that both $\cosh(x)$ and $I_0(x)$ are even functions, so $I^{(3)}$ 
and $I^{(4)}$ have an analytic power series expansion in powers of $X$.

We are interested in the matrix element in the limit $R \to \infty$.
Evaluating the matrix element in this limit therefore requires us to evaluate the integral
\[
\scr{I}^{(d)} &= \myint \frac{d^d p_i}{(2\pi)^d}\myint \frac{d^d p_f}{(2\pi)^d} \,
e^{-(\norm{p_i - p}^2 + \norm{p_f - p}^2)R^2 / 2}
I^{(d)}(p_i, p_f; y)
\nonumber\\
&= \myint \frac{d^d \bar{p}}{(2\pi)^d}\myint \frac{d^d q}{(2\pi)^d} \,
e^{-\norm{\bar{p} - p}^2 R^2} \, e^{-\norm{q}^2 R^2/4} \,
I^{(d)}(\bar{p} - \sfrac 12 q, \bar{p} + \sfrac 12 q; y).
\]
For large $R$, the integral is dominated by $q \simeq 0$
and $\bar{p} \simeq p$.
The $\bar{p}$ integral can be performed by saddle point integration,
because the saddle at $\bar{p} = p$ is in the region where all the step
functions involving $\bar{p}$ are nonzero.
This gives
\[
\scr{I}^{(d)} &= \left( \frac{\sqrt{\pi}}{2\pi R} \right)^d
\myint \frac{d^d q}{(2\pi)^d} \,
e^{-\norm{q}^2 R^2/4} \,
I^{(d)}(p - \sfrac 12 q, p + \sfrac 12 q; y).
\]
The integrand is proportional to $\th(-q^2)$, so the remaining $q$ integral
must be performed with care.
The integral can be evaluated in a reference frame where 
\[
\eql{rest}
p^\mu = (E, 0, \ldots, 0),
\qquad
E > 0.
\]
In this reference frame we have, to leading order in $q$, 
\[
D &= E^2 |\pvec{q}|^2,
\\
K^0 &= \sfrac 12 E,
\\
\vec{K} &= -\sfrac 12 E \la \hat{q},
\\
Y^0 &= y^0 - \vec{y} \cdot \hat{q},
\\
\vec{Y} &= \vec{y} - y^0 \hat{q},
\\
X &= \sfrac 14 (1 - \la^2) E^2 |\vec{y}_\perp|^2,
\]
where 
\[
\hat{q} = \frac{\vec{q}}{|\pvec{q}|},
\qquad
\la = \frac{q^0}{|\pvec{q}|},
\]
and $\vec{y}_\perp = \vec{y} - (\vec{y} \cdot \hat{q}) \hat{q}$
is the component of $\vec{y}$ that is perpendicular to $\vec{q}$. 
Expanding in the power series for either $I_0(x)$ or $\cosh (x)$ means we 
must evaluate the integrals
\[
\scr{J}^{(3)}_{\ell,m} &= \myint d^3 q\, \th(-q^2) \, e^{-\norm{q}^2 R^2/4} \,\frac{X^\ell (K \cdot y)^m}{2(-p_i^2 p_f^2 q^2)^{1/2}}\\
\scr{J}^{(4)}_{\ell,m} &= \myint d^4 q\, \th(-q^2) \, e^{-\norm{q}^2 R^2/4} \,
\frac{X^\ell (K \cdot y)^m}{8 D^{1/2}}
\]
We are not able to find a general formula for all $\ell$ and $m$,
but we are able to perform them analytically for the low values 
needed for this calculation.
We obtain
\begin{align}
\begin{split}
\mathcal{I}^{(4)}_0 &=C^{(4)}\left(\frac{E}{2}\right)^{-1},\\
\mathcal{I}^{(4)}_{1} &=C^{(4)} 
y_0,\\
\mathcal{I}^{(4)}_{2} &=C^{(4)}\frac{1}{6}\left(\frac{E}{2}\right) \left(|\vec{y}\gap|^2+3y_0^2\right),\\
\mathcal{I}^{(4)}_{3}&=C^{(4)}\frac{1}{6}\left(\frac{E}{2}\right)^{2}\left(|\vec{y}\gap|^2 +y_0^2\right)  y_0,\\
\mathcal{I}^{(4)}_{4} &=C^{(4)}\frac{1}{120}\left(\frac{E}{2}\right)^{3} \left(|\vec{y}\gap|^{4}+10y_0^2 |\vec{y}\gap|^2+5y_0^4\right),\\
\mathcal{I}^{(4)}_{5} &=C^{(4)}\frac{1}{360}\left(\frac{E}{2}\right)^{4} \left(3 |\vec{y}\gap|^4+10 y_0^2|\vec{y}\gap|^2+ 3y_0^4\right) y_0,\\
\mathcal{I}^{(4)}_{6} &=C^{(4)}\frac{1}{5040}\left(\frac{E}{2}\right)^{5} \left(|\vec{y}\gap|^{6}+21y_0^2|\vec{y}\gap|^4 +35y_0^4|\vec{y}\gap|^2 + 7 y_0^6\right),
\end{split}
\end{align} 
where $C^{(4)} = 1/{16(2\pi)^{9/2}R^7}$ and the subscript counts the number of powers of $y$. 
The index structures were combined by brute force using Mathematica,
and the details will not be presented here.

Following the same steps for $d = 3$ we obtain
\begin{align}
\begin{split}
\mathcal{I}^{(3)}_0 &=C^{(3)}\left(\frac{E}{2}\right)^{-2},\\
\mathcal{I}^{(3)}_{1} &=C^{(3)}\left(\frac{E}{2}\right)^{-1} 
y_0,\\
\mathcal{I}^{(3)}_{2} &=  C^{(3)}\frac{1}{4}\left(|\vec{y}\gap|^2+2y_0^2\right),\\
\mathcal{I}^{(3)}_{3}&= C^{(3)}\frac{1}{12}\left(\frac{E}{2}\right)^{1}\left(3|\vec{y}\gap|^2 + 2y_0^2\right)  y_0,\\
\mathcal{I}^{(3)}_{4} &=C^{(3)}\frac{1}{192}\left(\frac{E}{2}\right)^{2} \left(3|\vec{y}\gap|^{4}+24y_0^2 |\vec{y}\gap|^2+8y_0^4\right),\\
\mathcal{I}^{(3)}_{5} &= C^{(3)}\frac{1}{960}\left(\frac{E}{2}\right)^{3} \left(15 |\vec{y}\gap|^4+40 y_0^2|\vec{y}\gap|^2+ 8y_0^4\right) y_0,\\
\mathcal{I}^{(3)}_{6} &=  C^{(3)}\frac{1}{11520}\left(\frac{E}{2}\right)^{4} \left(5|\vec{y}\gap|^{6}+90y_0^2|\vec{y}\gap|^4 +120 y_0^4|\vec{y}\gap|^2 + 16 y_0^6\right),
\end{split}
\end{align}
where $C^{(3)} = 1/{32(2 \pi)^{5/2}R^5}$.
Differentiating these expressions appropriately with respect to $y$ and adding the necessary linear combations gives us the complete expression for the free scalar energy-momentum tensor 3-point function in both 3 and 4 dimensions. The result is presented in Appendix B.

\subsection{Free Fermions}

We follow the same procedure for the Dirac fermions. 
The energy-momentum tensor for a free Dirac fermion is given by 
\[
T_{\mu\nu} &= \frac{i}{2} \bar{\psi}\left(\gamma_\mu \!\! \stackrel{\leftrightarrow} \partial _\nu\!\!  + \gamma_\nu  \!\! \stackrel{\leftrightarrow}\partial _\mu\!\! \,\right)\psi,
\]
where 
\[
 \raisebox{1.5pt}{$\stackrel{\leftrightarrow}{\d}_\mu$}  
= \dfrac{1}{2}(\d_\mu -\raisebox{1.5pt}{$\stackrel{\leftarrow}{\d}_\mu$} ).
\]
We write the contraction of two fermion fields as 
\beq
\bcontraction{}{\psi}{^\al(x) }{\bar{\psi}}
\psi^\al(x) \bar{\psi}_\be(y)
= \myint \frac{d^d p}{(2\pi)^d}\, e^{-i p \cdot (x - y)} \, W^\al{}_\be(p),
\eeq
where the Wightman propagator is given by
\[
W^\al{}_\be(p)
= \th(p^0) 2\pi \de(p^2) (\sla{p})^\al{}_\be
= (\sla{p})^\al{}_\be W(p).
\]
To determine the 3-point function, we compute the 3-point function of the operator
\[
C_{\mu\nu} = \bar{\psi} \gamma_\mu \partial_\nu \psi- {\d_\nu \bar{\psi}} \ga_\mu \psi.
\]

All other relevant 3-point functions for the fermion case can be computed by 
permuting the indices on $\aavg{CCC}$.
The complete results are presented in Appendix B. As an example, we have
\[
\eql{resF}
&\aavg{\tilde{C}_{\mu\nu}(p_1) \tilde{C}_{\rho\si}(p_2) \tilde{C}_{\tau\om}(p_3)}
\nonumber\\
&\qquad\qquad{}
= 2i\Tr(\gamma_\alpha \gamma_\mu \gamma_\beta \gamma_\rho \gamma_\delta \gamma_\tau) \myint \frac{d^d k}{(2\pi)^d} \,
k^{\alpha} (p_f - k)^{\beta} (p_i-k)^\delta\nonumber\\
&\qquad\qquad\qquad\qquad\quad{}
\times \left[(p_f - k)_{\nu} -k_\nu\right]\left[(p_i - k)_{\si} + (p_f-k)_\si\right]\left[(p_i-k)_\om-k_\om\right]
\nonumber\\
&\qquad\qquad\qquad\qquad\quad{}
\times
W(k) W(p_f - k) W(p_i - k),
\]
where again $p_1 = p_f$ and $p_i = -p_3$. This has precisely the same form as the integrals that appear in the scalar case and can be computed using the same methods.

\subsection{Free Vectors}
A free vector can be described by a free $U(1)$ gauge theory.
The energy-momentum tensor is given by
\[
T_{\mu\nu} &= F_{\mu \lambda} F_{\nu}^{\ \lambda} - \frac{1}{4} \eta_{\mu\nu} F^{\rho\sigma} F_{\rho\sigma}.
\]
where $F_{\mu\nu} = \partial_\mu A_\nu - \partial_\nu A_\mu$ is the usual field strength tensor.
We write the contraction of two vector fields as 

\beq
\bcontraction{}{A}{_\mu(x) }{A}
A_\mu(x) A_\nu(y)
= \myint \frac{d^d p}{(2\pi)^d}\, e^{-i p \cdot (x - y)} \, W_{\mu\nu}(p), 
\eeq

where the Wightman propagator in Feynman gauge is given by
\[
W_{\mu\nu}(p) &= -\eta_{\mu\nu} \, 2\pi \de(p^2) \th(p^0) =  -\eta_{\mu\nu} \, W(p).
\]
Because $T^{\mu\nu}$ is a gauge invariant operator, the result does not
depend on this choice. The energy-momentum tensor 3-point function can be determined solely 
from the 3-point function of $\aavg{FFF}$ and its various traces. 
The results for the free vector look similar to \Eq{resS} and \Eq{resF}, although more complicated and not very illuminating (again full results are in Appendix B), so we can use the same procedure as in the scalar case to get the final expression.

\subsection{Expectation Value of the Energy-Momentum Tensor} 
After normalizing our matrix element with the two point function,
we obtain for the expectation value of $T^{\mu\nu}(0)$ in the state \Eq{wavepacket} 
\begin{align}
\begin{split}
\eql{Expect4}
\avg{T_{\mu\nu}}^{(4)} 
&= \frac{\sqrt{p^2}}{R^3} \al \biggl[
(12 n_v-3n_f - n_s )\frac{(\ep^*\ep)_{(\mu\nu)}}{\Tr(\ep^*\ep) }
\\
&\qquad\qquad\qquad{}
+(13n_f + 2n_s + 32 n_v)\frac{p_\mu p_\nu}{p^2}
\\
&\qquad\qquad\qquad{}
-\frac{1}{4}(44 n_v+ 10n_f + n_s )\eta_{\mu\nu}  \biggr]
\end{split}
\end{align}
and 
\begin{align}
\eql{Expect3}
\begin{split}
\langle T_{\mu\nu} \rangle^{(3)} 
&=\ \frac{\sqrt{p^2}}{R^2}\beta\biggl[ -4(n_f +n_s)\frac{(\ep^*\ep)_{(\mu\nu)}}{\Tr(\ep^*\ep) }
\\
&\qquad\qquad\qquad{}
+ (7n_s + 16 n_f)\frac{p_\mu p_\nu}{p^2}
\\
&\qquad\qquad\qquad{}
-(n_s + 4 n_f) \eta_{\mu\nu}  \biggr],
\end{split}
\end{align}
where $\alpha$ and $\beta$ are defined in \Eqs{alpha} and \eq{beta}, respectively. 
Although the normalization of the states is not needed for our bounds,
we computed it to check that we reproduce the known dependence on $n_s$, $n_f$, and $n_v$.
These expressions are then used in section 2 to find our constraints.

This concludes our general discussion of our calculations.
Further details can be found in Appendix B.

\section{Conclusions}
We have shown that states in a 4D CFT defined by acting with the energy-momentum tensor
on the vacuum can have an expectation value for the spacetime averaged energy density 
that is arbitrarily negative, unless the $TTT$ OPE coefficients obey certain
inequalities specified in \Eq{ourconstraints}.
If these inequalities are violated, the theory has 
states that are indistinguishable from states
of constant negative energy density by any local measurement.
This strongly
suggests that the theory suffers from instabilities or other inconsistencies.
The constraints we find are weaker than the inequalities of Hofman and Maldacena,
but we argue that there may be consistent theories that violate these inequalities.

The calculations in this paper were carried out by computing Wightman correlation
functions directly in momentum space.
We believe that these methods may have some interest independently of the
results above.

There are a number of directions for future work.
It is straightforward to extend it to states created by other operators, but that will
not shed light on the conceptual questions raised in this work.
It would be very interesting to understand how to probe the existence of
instabilities in an abstract conformal field theory, defined by an operator
spectrum and OPE coefficients.
If an instability can be exhibited for some range of OPE coefficients,
that would put these bounds on a firm foundation. 
Another direction would be to constrain the $TTT$ OPE coefficients directly
from unitarity and crossing symmetry using the conformal bootstrap.
It would be interesting to know whether the constraints we are describing
are already implied by the crossing equations.

\section*{Acknowledgements}
We thank R. Rattazzi, J. Thaler, and A. Waldron for discussions.
M.A.L. wishes to thank the Kavli Institute for Theoretical Physics in
Santa Barbara, the Aspen Center for Physics, and the Galileo Galilei Institute 
for hospitality during this work.
This work was supported by DOE
grant DE-FG02-91ER40674.

\appendix{Appendix A: Null Versus Weak Energy Condition in CFT}
The null energy condition (NEC) is the statement that
for any null vector $n^\mu$ ($n^2 = 0$)
we have $T^{\mu\nu} n_\mu n_\nu \ge 0$.
The weak energy condition (WEC) is the statement that for
any timelike vector $u^\mu$ we have $T^{\mu\nu} u_\mu u_\nu \ge 0$.
It is easy to see that if the NEC is violated, then so is the WEC.
The reason is simply that in a general reference frame we can write
\beq
n^\mu = (1, \hat{n}),
\eeq
where $\hat{n}$ is a unit vector,
and we can choose
\beq
u^\mu = \bigl(\cosh(y),\ \sinh(y) \gap \hat{n}\bigr).
\eeq
This is timelike; in fact it satisfies $u^2 = 1$.
For large $y$, $u^\mu \to \cosh(y)\gap n^\mu$ and we have
\beq
T^{\mu\nu} u_\mu u_\nu \to \cosh^2(y) \gap T^{\mu\nu} n_\mu n_\nu < 0.
\eeq

We can ask whether there is a converse to this statement:
does violation of the WEC imply violation of the NEC?
A simple counterexample is given by a negative cosmological constant,
$T^{\mu\nu} = -\Lambda^4 \eta^{\mu\nu}$,
which has $T^{\mu\nu} u_\mu u_\nu < 0$ and $T^{\mu\nu} n_\mu n_\nu = 0$ 
for all $u^\mu$ and $n^\mu$.
However, this is in some sense the only counterexample.
Specifically, we show that if we restrict to energy-momentum tensors that are
traceless (as in a CFT), then violation of the WEC implies violation of the
NEC. 

We work in a reference frame where the energy-momentum
tensor is diagonal:
\beq
T^{\mu\nu} = \diag(\rho,\, -p_1,\, \ldots,\, -p_{d-1}).
\eeq
We have defined the components so that $p_i$ is the pressure
density in the $i$ direction.
Tracelessness of the energy-momentum tensor then implies
\beq\eql{tracelessness}
\rho = -\sum_i p_i.
\eeq

The reference frame where the energy density is negative can be defined by
the velocity vector
\beq
u^\mu = (\cosh(y),\ \sinh(y) \hat{u}),
\eeq
where $\hat{u}$ is a unit vector.
Then we have
\beq\eql{negativeenergy}
0 > T^{\mu\nu} u_\mu u_\nu = \cosh^2(y) \biggl[ 
\rho - \tanh^2(y) \sum_i \hat{u}_i^2 p_i \biggr].
\eeq
The question is whether we can find a null vector 
\beq
n^\mu = (1,\ \hat{n})
\eeq
so that
\beq\eql{NECviolation}
0 \stackrel{?}> T^{\mu\nu} n_\mu n_\nu
= \rho - \sum_i \hat{n}_i^2 p_i.
\eeq
Note that $0 \le \tanh^2(y) < 1$.
Therefore, if $\rho \ge 0$, then \Eq{negativeenergy} can be satisfied only if
$\sum_i \hat{u}_i^2 p_i > \rho$.
But then we can satisfy \Eq{NECviolation} by simply taking
$\hat{n} = \hat{u}$.
If $\rho < 0$, we use the fact that tracelessness condition \Eq{tracelessness}
implies that at least one of the $p_i$ is strictly positive.
We can then simply choose $\hat{n}$ to point in the direction of one
of the positive $p_i$ to satisfy \Eq{NECviolation}.

\appendix{Appendix B: Momentum Space Wightman Functions}

The following are the results of the calculation of the momentum space Wightman functions for the 2- and 3-point functions of the energy-momentum tensor for a free scalar, free fermion and free vector field in both 3D and 4D. The expressions found here are true in general, although approximations were applied to derive the bounds in section 2.

\subsection{3-Point Function Tensor Structures}
We write the energy-momentum tensor 3-point function
\begin{align}\eql{TTT}
\aavg{ T^{\mu\nu}(p_f) T^{\rho\sigma}(-q) T^{\tau\omega}(-p_i)}
\end{align}
as a sum of products of three different tensor structures, $\mathcal{S}^{\alpha\beta},\ \mathcal{F}^{\alpha\beta\gamma},\ \mathcal{H}^{\mu_1\cdots \mu_n}$ (defined below) and their contractions. 
Since we are concentrating on the 3-point function for the energy-momentum tensor, every term must have a 
total of 6 indices. 
We define
\begin{align}
p^\alpha_A &= \left\{p^\alpha_1, p^\alpha_2, p^\alpha_3\right\}= \left\{p_f^\alpha, -q^\alpha, -p_i^\alpha\right\},
\end{align}
where $A$ labels the momentum.

The simplest tensor structure that can appear is 
\begin{align}
\mathcal{S}^{\, \alpha\, \beta}_{AB} (p) =  p_{A}^{\, \beta}p_{B}^{\, \alpha}-\left(p_{A}\cdot p_{B}\right) \eta^{\alpha\beta}
\end{align}
where
$\mathcal{S}$ is obviously conserved: $p_{A \alpha} \mathcal{S}^{\, \alpha\, \beta}_{AB} (p)= p_{B \beta} \mathcal{S}^{\, \alpha\, \beta}_{AB} (p) = 0$.

The second tensor structure is a linear combination of the first, and is defined only for specific momenta (and their corresponding Lorentz) indices:
    \begin{align}
  \mathcal{F}^{\alpha\beta\gamma}_{123}(p) 
&=\  \mathcal{S}^{\alpha\beta}_{13} p_2^\gamma +\mathcal{S}^{\alpha\gamma}_{32} p_1^\beta +\mathcal{S}^{\beta\gamma}_{12} p_3^\alpha-\mathcal{S}^{\alpha\beta}_{23} p_1^\gamma -\mathcal{S}^{\alpha\gamma}_{12} p_3^\beta -\mathcal{S}^{\beta\gamma}_{31} p_2^\alpha 
  \end{align}
   This means that $\alpha$ must be an index on the left energy-momentum tensor (since it corresponds with a momentum index of 1), $\beta$ must be an index on the middle energy-momentum tensor, and $\gamma$ must be an index on the right energy-momentum tensor in the 3-point function. 
The contractions of this tensor structure also appear:
\begin{align}
\mathcal{F}^{\alpha\beta} &=\ -p_{3\lambda}\mathcal{F}^{\lambda\alpha\beta}_{123} = p_{1\lambda} \mathcal{F}^{\alpha\beta\lambda}_{123}  \nonumber\\
&=\ D \eta^{\alpha\beta} -\frac{1}{2}\left( p_1^2 p_2^\alpha p_3^\beta + p_2^2 p_1^\alpha p_3^\beta + p_3^2 p_1^\alpha p_2^\beta\right)
\end{align}
and 
\begin{align}
\mathcal{F}^\alpha_1 = \eta_{\beta\gamma} \mathcal{F}^{\alpha\beta\gamma}_{123},\ \ \ \ \ \mathcal{F}^\beta_2 = \eta_{\alpha\gamma} \mathcal{F}^{\alpha\beta\gamma}_{123},\ \ \ \ \ \mathcal{F}^\gamma_3 = \eta_{\alpha\beta} \mathcal{F}^{\alpha\beta\gamma}_{123}
\end{align}
Note that the 2-index structure $\mathcal{F}^{\alpha\beta}$ has no momentum indices, since it is defined the same way for any momentum combination.

The final tensor structure $\mathcal{H}$ can have any number (1 to 6) of indices and is completely symmetric.  With one index, 
\begin{align}
\mathcal{H}^{\alpha}_{A} &= 2I^{\alpha} - \ell_{A}^{\alpha} I^0
\end{align}
where, as defined in the main text, 
\begin{align}
I^0 &= \int \frac{d^dk}{(2\pi)^d} W(k) W(p_f-k)W(p_i-k), \\
I^{\mu_1\cdots \mu_n} &= \int \frac{d^dk}{(2\pi)^d}k^{\mu_1}\cdots k^{\mu_n} W(k) W(p_f-k)W(p_i-k)\eql{sym}
\end{align}
and 
\begin{align}
\ell^\alpha_A &= \left\{\ell^\alpha_1, \ell^\alpha_2, \ell^\alpha_3\right\}= \left\{p_f^\alpha, p_i^\alpha+p_f^\alpha, p_i^\alpha\right\}.
\end{align}

With two indices, this tensor structure becomes
\begin{align}
\mathcal{H}^{\, \alpha\, \beta}_{AB} &= 2I^{\alpha\beta} - \ell_{A}^{\alpha} I^{\beta}- \ell_{B}^{\beta} I^{\alpha} + \frac{1}{2} \ell_{A}^{\alpha}\ell_{B}^{\beta}I^0.
\end{align}
The pattern continues for higher indices. The first term in the sum is always $I$ with the maximum number of indices and a coefficient of $2$. Each succeeding term has one more power of $\ell$ and one fewer index on $I$. The coefficients decrease by a factor of two and alternate in sign. For example, with six indices,
\begin{align}
\mathcal{H}^{\, \alpha\, \beta\, \gamma\, \delta\, \epsilon\, \zeta}_{ABCDEF} &= 2I^{\alpha\beta\gamma\delta\epsilon\zeta} \\
&- \ell_{A}^{\alpha} I^{\beta\gamma\delta\epsilon\zeta}- \ell_{B}^{\beta} I^{\alpha\gamma\delta\epsilon\zeta}- \ell_{C}^{\gamma} I^{\alpha\beta\delta\epsilon\zeta}- \ell_{D}^{\delta} I^{\alpha\beta\gamma\epsilon\zeta}- \ell_{E}^{\epsilon} I^{\alpha\beta\gamma\delta\zeta}- \ell_{F}^{\zeta} I^{\alpha\beta\gamma\delta\epsilon}\nonumber\\
&+\cdots \nonumber\\
&-\frac{1}{16} \ell_{B}^{\beta}\ell_{C}^{\gamma}\ell_{D}^{\delta}\ell_{E}^{\epsilon}\ell_{F}^{\zeta}I^\alpha - \frac{1}{16}\ell_{A}^{\alpha}\ell_{C}^{\gamma}\ell_{D}^{\delta}\ell_{E}^{\epsilon}\ell_{F}^{\zeta}I^\beta- \frac{1}{16}\ell_{A}^{\alpha}\ell_{B}^{\beta}\ell_{D}^{\delta}\ell_{E}^{\epsilon}\ell_{F}^{\zeta}I^\gamma\nonumber\\
&- \frac{1}{16}\ell_{A}^{\alpha}\ell_{B}^{\beta}\ell_{C}^{\gamma}\ell_{E}^{\epsilon}\ell_{F}^{\zeta}I^\delta- \frac{1}{16}\ell_{A}^{\alpha}\ell_{B}^{\beta}\ell_{C}^{\gamma}\ell_{D}^{\delta}\ell_{F}^{\zeta}I^\epsilon- \frac{1}{16}\ell_{A}^{\alpha}\ell_{B}^{\beta}\ell_{C}^{\gamma}\ell_{D}^{\delta}\ell_{E}^{\epsilon}I^\zeta\nonumber\\
&+\frac{1}{32} \ell_{A}^{\alpha}\ell_{B}^{\beta}\ell_{C}^{\gamma}\ell_{D}^{\delta}\ell_{E}^{\epsilon}\ell_{F}^{\zeta}I^0\nonumber.
\end{align}

\subsection{2-Point Function Tensor Structures}
The energy-momentum tensor 2-point function
\begin{align}\eql{TT}
\aavg{T^{\mu\nu}(p) T^{\rho\sigma}(-p) }
\end{align}
can be described using similar notation, with some obvious restrictions. 
The momentum, before $\ell_A^\alpha$ and $p_A^\alpha$, now only has one possibility, $p^\alpha$, so we drop the momentum index. 

Restricted to describe the 2-point function, the first tensor structure $\mathcal{R}$, defined similarly to the 3-point structure $\mathcal{S}$, is:
\begin{align}
\mathcal{R}^{ \alpha\beta}  =  p^{\alpha}p^{\beta}-p^2 \eta^{\alpha\beta}
\end{align}
which again is obviously conserved.

The second tensor structure is defined by the same pattern as $\mathcal{H}$ for the 3-point function, but with the necessary restrictions on momenta. For example:
\begin{align}
\mathcal{G}^{\alpha} &= 2Y^{\alpha} - p^{\alpha} Y^0\\
\mathcal{G}^{\alpha\beta} &= 2Y^{\alpha\beta} - p^{\alpha} Y^{\beta}- p^{\beta} Y^{\alpha} + \frac{1}{2} p^{\alpha}p^{\beta}Y^0
\end{align}
where, similar to the 3-point function's $I^0$ and $I^{\mu_1\cdots \mu_n}$, 
\begin{align}
Y^0 &= \int \frac{d^dk}{(2\pi)^d} W(k) W(p-k)\\
Y^{\mu_1\cdots \mu_n} &= \int \frac{d^dk}{(2\pi)^d}k^{\mu_1}\cdots k^{\mu_n} W(k) W(p-k).
\end{align}

\subsubsection{Free Scalar}

For a free scalar, the energy-momentum tensor can be written
\begin{align}
T^{\alpha\beta}_{s} =A^{\alpha\beta}- \frac{1}{2} \eta^{\alpha\beta} \eta_{\kappa\lambda} A^{\kappa\lambda} -\frac{1}{2}\xi\left( \eta^{\alpha\beta} \eta_{\kappa\lambda} B^{\kappa\lambda} - B^{\alpha\beta}\right),
\end{align}
where $A^{\alpha\beta} = \partial^\alpha \phi \partial^\beta\phi$, $B^{\alpha\beta} = \partial^\alpha \partial^\beta (\phi^2)$ and $\xi = \frac{d-2}{2(d-1)}$ in a conformal theory. The full 2-point function will be a linear combination of 2-point functions of the $A$'s and $B$'s. These are:

\begin{align}
\aavg{B^{\mu\nu}(p) B^{\rho\sigma} (-p)} &=\ 2 p^{\mu} p^{\nu} p^{\rho} p^{\sigma} Y^0 \\
\aavg{B^{\mu\nu}(p) A^{\rho\sigma}(-p) }&=\  - p^{\mu} p^{\nu} \mathcal{G}^{\rho\sigma} + \frac{1}{2}p^{\mu} p^{\nu} p^{\rho} p^{\sigma} Y^0\\
\aavg{A^{\mu\nu}(p) B^{\rho\sigma}(-p) } &=\  - p^{\rho} p^{\sigma} \mathcal{G}^{\mu\nu}  + \frac{1}{2}p^{\mu} p^{\nu} p^{\rho} p^{\sigma} Y^0\\
\aavg{A^{\mu\nu}(p)A^{\rho\sigma}(-p) } &=\  \mathcal{G}^{\mu\nu\rho\sigma} -\frac{1}{4} \left( p^\mu p^\nu \mathcal{G}^{\rho\sigma}  + p^\rho p^\sigma \mathcal{G}^{\mu\nu}\right)+\frac{1}{8} p^\mu p^\nu p^\rho p^\sigma Y^0.
\end{align}

Combining these into full 2-point function for the free scalar gives:
\begin{align}
\aavg{T^{\mu\nu} (p)T^{\rho\sigma}(-p)}_{\text{scalar}}^{(d)} &=\ \mathcal{G}^{\mu\nu\rho\sigma} +\frac{1}{2}\left(\xi-\tfrac{1}{2}\right)\left(\mathcal{R}^{\rho\sigma} \mathcal{G}^{\mu  \nu }+\mathcal{R}^{\mu\nu} \mathcal{G}^{\rho  \sigma }\right) \nonumber\\
&+\frac{1}{2}\left(\xi-\tfrac{1}{2}\right)^2\mathcal{R}^{\mu\nu}\mathcal{R}^{\rho\sigma}Y^0
\end{align}

\subsubsection{Free Dirac Fermion}
For a free Dirac fermion we have
\begin{align}\eql{Tfermion}
T^{\alpha\beta}_f = \frac{i}{2} \left( C^{\alpha\beta} + C^{\beta\alpha}\right)
\end{align}
with $C^{\alpha\beta} = \bar{\psi} \gamma^\alpha\partial^\beta \psi- \partial^\beta \bar{\psi} \gamma^\alpha \psi$.

For one particular term in our 2-point function, we get
\begin{align}
\left(\tfrac{i}{2}\right)^2\aavg{C^{\mu\nu}C^{\rho\sigma} }^{(d)} &=\ \frac{1}{4}\Tr(\gamma_d)\left( \mathcal{G}^{\mu  \nu  \rho  \sigma }+\frac{1}{4}\mathcal{R}^{\mu\rho} \mathcal{G}^{\nu  \sigma }\right)\eql{CC}
\end{align}
where $\Tr(\gamma_d)$ is the trace of the identity matrix in the dimension of the representation of the gamma matrices:
\begin{align}
\Tr(\gamma_d) &= 2^{d/2} \textnormal{ for $d$ = even}\\
\Tr(\gamma_d) &= 2^{(d-1)/2} \textnormal{ for $d$ = odd}.
\end{align}
So in $d = 3$ we have $\Tr(\gamma_d) = 2$, while in $d = 4$, $\Tr(\gamma_d) = 4$. The full 2-point function is then found by permuting $\mu\leftrightarrow \nu$ and $\rho\leftrightarrow \sigma$ in 
\Eq{CC} and adding the results:
\begin{align}
\aavg{T^{\mu\nu} T^{\rho\sigma}}^{(d)}_{\text{fermion}} &=\ \Tr(\gamma_d)\mathcal{G}^{\mu  \nu  \rho  \sigma }\nonumber\\
&+ \frac{1}{16}\Tr(\gamma_d)\left(\mathcal{R}^{\mu\rho} \mathcal{G}^{\nu  \sigma }+\mathcal{R}^{\nu\rho} \mathcal{G}^{\mu  \sigma }+\mathcal{R}^{\mu\sigma} \mathcal{G}^{\nu  \rho }+\mathcal{R}^{\nu\sigma} \mathcal{G}^{\mu  \rho }\right).
\end{align}

\subsubsection{Free Vector}

For a free vector field we have
\begin{align}\eql{Tvector}
T^{\alpha\beta}_{v} = F^{\alpha \lambda} F^{\beta}_{\ \lambda} - \frac{1}{4} \eta^{\alpha\beta} F^{\kappa\lambda} F_{\kappa\lambda}
\end{align}
where $F^{\alpha\beta}= \partial^\alpha A^\beta - \partial^\beta A^\alpha $ is the usual field strength tensor. The free vector is only conformal in $d = 4$, so the following calculations are 4D-specific. For the 2-point function, the first term is
\begin{align}
\aavg{F^{\mu \lambda} F^{\nu}_{\ \lambda}(p) F^{\rho\kappa} F^{\sigma}_{\ \kappa}(-p)} &=\ 2\mathcal{G}^{\mu  \nu  \rho  \sigma }\\
&+\frac{1}{2}\left( \mathcal{R}^{\rho\sigma} \mathcal{G}^{\mu\nu} + \mathcal{R}^{\mu\nu} \mathcal{G}^{\rho\sigma}  \right)\nonumber\\
&-\frac{1}{2}\left( \mathcal{R}^{\mu\rho} \mathcal{G}^{\nu\sigma} + \mathcal{R}^{\mu\sigma} \mathcal{G}^{\nu\rho} + \mathcal{R}^{\nu\rho} \mathcal{G}^{\mu\sigma} + \mathcal{R}^{\nu\sigma} \mathcal{G}^{\mu\rho} \right)\nonumber\\
&+\frac{1}{4}\left( \mathcal{R}^{\mu\rho} \mathcal{R}^{\nu\sigma} + \mathcal{R}^{\nu\rho} \mathcal{R}^{\mu\sigma}\right)Y^0\nonumber\\
&+\frac{1}{4}\left(p^2 \eta^{\mu\nu} p^\rho p^\sigma + p^2 \eta^{\rho\sigma} p^\mu p^\nu -p^\mu p^\nu p^\rho p^\sigma\right) Y^0\nonumber
\end{align}

The full 2-point function is calulated by taking various traces of the first term and combining them, 

\begin{align}
\aavg{T^{\mu\nu}(p) T^{\rho\sigma}(-p)}^{(4)}_{\text{vector}} &=\ 2\mathcal{G}^{\mu\nu\rho\sigma}\\
&+\frac{1}{2}\left( \mathcal{R}^{\rho\sigma} \mathcal{G}^{\mu\nu} + \mathcal{R}^{\mu\nu} \mathcal{G}^{\rho\sigma}  \right)\nonumber\\
&-\frac{1}{2}\left(  \mathcal{R}^{\mu\rho} \mathcal{G}^{\nu\sigma} + \mathcal{R}^{\mu\sigma} \mathcal{G}^{\nu\rho} + \mathcal{R}^{\nu\rho} \mathcal{G}^{\mu\sigma} + \mathcal{R}^{\nu\sigma} \mathcal{G}^{\mu\rho} \right)\nonumber\\
&+\frac{1}{4}\left( \mathcal{R}^{\mu\rho} \mathcal{R}^{\nu\sigma} + \mathcal{R}^{\nu\rho} \mathcal{R}^{\mu\sigma}-\mathcal{R}^{\mu\nu}\mathcal{R}^{\rho\sigma}\right)Y^0\nonumber
\end{align}

\subsection{3-point Functions}

\subsubsection{Free Scalar}

For the 3-point function of the free scalar, using the definition in \Eq{Tscalar}, we get:
 \begin{align}
 \aavg{B^{\mu\nu}(p_1) B^{\rho\sigma}(p_2) B^{\tau\omega}(p_3) } &=\  -8 p_1^{\mu} p_1^{\nu} p_2^\rho p_2^\sigma p_3^\tau p_3^\omega I^0\\
 \aavg{A^{\mu\nu}(p_1) B^{\rho\sigma}(p_2) B^{\tau\omega}(p_3) } &=\ 4 p_2^\rho p_2^\sigma p_3^\tau p_3^\omega \mathcal{H}_{11}^{\mu\nu} -2 p_1^{\mu} p_1^{\nu} p_2^\rho p_2^\sigma  p_3^\tau p_3^\omega I^0\nonumber \\
  \aavg{B^{\mu\nu} (p_1)A^{\rho\sigma}(p_2) B^{\tau\omega}(p_3) } &=\  4  p_1^{\mu} p_1^{\nu} p_3^\tau p_3^\omega \mathcal{H}_{22}^{\rho\sigma} -2 p_1^{\mu} p_1^{\nu}p_2^\rho p_2^\sigma  p_3^\tau p_3^\omega I^0\nonumber \\
 \aavg{B^{\mu\nu}(p_1) B^{\rho\sigma}(p_2) A^{\tau\omega}(p_3) } &=\  4  p_1^{\mu} p_1^{\nu} p_2^\rho p_2^\sigma  \mathcal{H}_{33}^{\tau\omega} -2 p_1^{\mu} p_1^{\nu} p_2^\rho p_2^\sigma p_3^\tau p_3^\omega I^0 \nonumber\\
\aavg{B^{\mu\nu}(p_1) A^{\rho\sigma}(p_2) A^{\tau\omega} (p_3)}&=\ -4 p_1^{\mu} p_1^{\nu}\mathcal{H}_{2233}^{\rho\sigma\tau\omega} +  p_1^{\mu} p_1^{\nu}p_2^\rho p_2^\sigma \mathcal{H}_{33}^{\tau\omega} +  p_1^{\mu} p_1^{\nu} p_3^\tau p_3^\omega \mathcal{H}_{22}^{\rho\sigma}\nonumber\\
&\quad\quad-\tfrac{1}{2}  p_1^{\mu} p_1^{\nu}p_2^\rho p_2^\sigma  p_3^\tau p_3^\omega I^0\nonumber\\
\aavg{A^{\mu\nu}(p_1) A^{\rho\sigma}(p_2) B^{\tau\omega}(p_3) } &=\  -4p_3^\tau p_3^\omega\mathcal{H}_{1122}^{\mu\nu\rho\sigma} +  p_2^\rho p_2^\sigma p_3^\tau p_3^\omega \mathcal{H}_{11}^{\mu\nu} +  p_1^{\mu} p_1^{\nu} p_3^\tau p_3^\omega\mathcal{H}_{22}^{\rho\sigma}\nonumber\\
&\quad\quad-\tfrac{1}{2}  p_1^{\mu} p_1^{\nu}p_2^\rho p_2^\sigma  p_3^\tau p_3^\omega I^0\nonumber\\
\aavg{A^{\mu\nu}(p_1) B^{\rho\sigma}(p_2) A^{\tau\omega} (p_3)}&=\  -4p_2^\rho p_2^\sigma\mathcal{H}_{1133}^{\mu\nu\tau\omega}+  p_1^{\mu} p_1^{\nu}p_2^\rho p_2^\sigma\mathcal{H}_{33}^{\tau\omega} + p_2^\rho p_2^\sigma p_3^\tau p_3^\omega \mathcal{H}_{11}^{\mu\nu}\nonumber\\
&\quad\quad-\tfrac{1}{2}   p_1^{\mu} p_1^{\nu} p_2^\rho p_2^\sigma p_3^\tau p_3^\omega I^0\nonumber\\
\aavg{A^{\mu\nu}(p_1) A^{\rho\sigma}(p_2) A^{\tau\omega}(p_3) } &=\ 4 \mathcal{H}_{112233}^{\mu\nu\rho\sigma\tau\omega}-  p_1^{\mu} p_1^{\nu} \mathcal{H}_{2233}^{\rho\sigma\tau\omega}-p_3^\tau p_3^\omega \mathcal{H}_{1122}^{\mu\nu\rho\sigma}-p_2^\rho p_2^\sigma \mathcal{H}_{1133}^{\mu\nu\tau\omega}\nonumber\\
&\quad\quad+\tfrac{1}{4}  p_1^{\mu} p_1^{\nu}p_2^\rho p_2^\sigma \mathcal{H}_{33}^{\tau\omega}+\tfrac{1}{4} p_2^\rho p_2^\sigma p_3^\tau p_3^\omega \mathcal{H}_{11}^{\mu\nu}+\tfrac{1}{4}  p_1^{\mu} p_1^{\nu} p_3^\tau p_3^\omega \mathcal{H}_{22}^{\rho\sigma} \nonumber\\
&\quad\quad- \tfrac{1}{8} p_1^{\mu} p_1^{\nu} p_2^\rho p_2^\sigma  p_3^\tau p_3^\omega I^0\nonumber
\end{align}
We have fixed the index structure here to be $\aavg{T^{\mu\nu}(p_1) T^{\rho\sigma}(p_2) T^{\tau\omega}(p_3)}$, so the momentum corresponding to $\mu,\ \nu$ is 1, the momentum corresponding to $\rho,\ \sigma$ is 2 and the momentum corresponding to $\tau,\ \omega$ is 3. We fix this index order for the 3-point function from now on, and so can drop the momentum indices to avoid clutter.

 Combining these into the full 3-point function for the free scalar gives:
\begin{align}
\aavg{T^{\mu\nu}(p_1) T^{\rho\sigma}(p_2) T^{\tau\omega}(p_3)}_{\text{scalar}}^{(d)}&=\ 4 \mathcal{H}^{\mu\nu\rho\sigma\tau\omega}\\
&+2 \left(\xi-\tfrac{1}{2}\right)\left( \mathcal{S}^{\rho\sigma}\mathcal{H}^{\mu\nu\tau\omega}+\mathcal{S}^{\mu\nu}\mathcal{H}^{\rho\sigma\tau\omega} +\mathcal{S}^{\tau\omega}\mathcal{H}^{\mu\nu\rho\sigma}\right)\nonumber\\
&+(\xi-\tfrac{1}{2})^2\left(\mathcal{S}^{\mu\nu} \mathcal{S}^{\rho\sigma} \mathcal{H}^{\tau\omega}+\mathcal{S}^{\mu\nu}\mathcal{S}^{\tau\omega}\mathcal{H}^{\rho\sigma}+\mathcal{S}^{\rho\sigma}\mathcal{S}^{\tau\omega}\mathcal{H}^{\mu\nu}\right)\nonumber\\
 &+(\xi-\tfrac{1}{2})^3\mathcal{S}^{\mu\nu}\mathcal{S}^{\rho\sigma}\mathcal{S}^{\tau\omega}I^0\nonumber
\end{align}

\subsubsection{Free Dirac Fermion}
For the 3-point function of the free Dirac fermion, using the definition in \Eq{Tfermion}
the result for one particular term is
\begin{align}\eql{CCC}
\left(\frac{i}{2}\right)^3\aavg{C^{\mu\nu} C^{\rho\sigma} C^{\tau\omega}} &=\ -\frac{1}{2} \Tr(\gamma_d)\mathcal{H}^{\mu\nu\rho\sigma\tau\omega}\nonumber\\
&-\frac{1}{8}\Tr(\gamma_d)\left(\mathcal{H}^{\nu\sigma\tau\omega}\mathcal{S}^{\mu\rho}+\mathcal{H}^{\nu\rho\sigma\omega}\mathcal{S}^{\mu\tau}+\mathcal{H}^{\mu\nu\sigma\omega}\mathcal{S}^{\rho\tau} \right)\nonumber\\
&+\frac{1}{16}\Tr(\gamma_d)\mathcal{H}^{\nu\sigma\omega} \mathcal{F}^{\mu\rho\tau}
\end{align}
where again we have fixed the index structure as in the scalar case and dropped the momentum indices for clarity.

We can get the complete expression for the energy-momentum tensor 3-point function of the free fermion by permuting $\mu\leftrightarrow \nu$, $\rho\leftrightarrow \sigma$ and $\tau\leftrightarrow \omega$ in \Eq{CCC} and adding the results:
\begin{align}
 \aavg{T^{\mu\nu} T^{\rho\sigma} T^{\tau\omega}}_{\text{fermion}}^{(d)}
 &=\  -4\Tr(\gamma_d)\mathcal{H}^{\mu  \nu  \rho  \sigma  \tau  \omega }\\
 &-\frac{1}{4}\Tr(\gamma_d)\bigg(\mathcal{S}^{\sigma  \omega } \mathcal{H}^{\mu  \nu  \rho  \tau }+\mathcal{S}^{\rho  \omega }\mathcal{H}^{\mu  \nu  \sigma \tau }+\mathcal{S}^{\sigma  \tau } \mathcal{H}^{\mu  \nu  \rho  \omega } +\mathcal{S}^{\rho  \tau }\mathcal{H}^{\mu  \nu  \sigma  \omega }\nonumber \\
&\quad\quad\quad\quad\quad   +\mathcal{S}^{\nu  \omega } \mathcal{H}^{\mu  \rho  \sigma  \tau }+ \mathcal{S}^{\mu  \omega } \mathcal{H}^{\nu  \rho  \sigma  \tau } +\mathcal{S}^{\nu  \tau }\mathcal{H}^{\mu  \rho  \sigma  \omega } +\mathcal{S}^{\mu  \tau }\mathcal{H}^{\nu  \rho  \sigma  \omega } \nonumber\\
&\quad\quad\quad\quad\quad  +\mathcal{S}^{\nu  \sigma } \mathcal{H}^{\mu  \rho  \tau  \omega }+ \mathcal{S}^{\mu \sigma } \mathcal{H}^{\nu  \rho  \tau  \omega }+\mathcal{S}^{\nu  \rho }\mathcal{H}^{\mu  \sigma  \tau  \omega } +\mathcal{S}^{\mu  \rho }\mathcal{H}^{\nu  \sigma  \tau  \omega }\bigg)\nonumber\\
&+ \frac{1}{16} \Tr(\gamma_d)\bigg(\mathcal{H}^{\mu  \rho  \tau}\mathcal{F}^{\nu\sigma\omega}+\mathcal{H}^{\nu  \rho  \tau}\mathcal{F}^{\mu\sigma\omega} +\mathcal{H}^{\mu  \sigma  \tau}\mathcal{F}^{\nu\rho\omega}+\mathcal{H}^{\nu  \sigma  \tau}\mathcal{F}^{\mu\rho\omega}\nonumber\\
&\quad\quad\quad\quad\quad +\mathcal{H}^{\mu  \rho  \omega}\mathcal{F}^{\nu\sigma\tau}+\mathcal{H}^{\nu  \rho  \omega}\mathcal{F}^{\mu\sigma\tau} +\mathcal{H}^{\mu  \sigma  \omega}\mathcal{F}^{\nu\rho\tau}+\mathcal{H}^{\nu  \sigma  \omega}\mathcal{F}^{\mu\rho\tau}\bigg).\nonumber
\end{align}

\subsection{Free Vector}

For the 3-point function of the free vector, using \Eq{Tvector}, the first term is:
\begin{align}
\aavg{F^{\mu\lambda}F^\nu_{\ \lambda}(p_1) F^{\rho\kappa}&F^\sigma_{\ \kappa}(p_2)F^{\tau\alpha}F^\omega_{\ \alpha}(p_3)}\nonumber\\
 &=\ 8\mathcal{H}^{\mu\nu\rho\sigma\tau\omega}\\
&+2\mathcal{S}^{\mu\nu} \mathcal{H}^{\rho\sigma\tau\omega}+2\mathcal{S}^{\rho\sigma}\mathcal{H}^{\mu\nu\tau\omega}+2\mathcal{S}^{\tau\omega}\mathcal{H}^{\mu\nu\rho\sigma}\nonumber\\
&+2\mathcal{S}^{\mu\rho}\mathcal{H}^{\nu\sigma\tau\omega}+2\mathcal{S}^{\nu\rho} \mathcal{H}^{\mu\sigma\tau\omega}+2\mathcal{S}^{\mu\sigma} \mathcal{H}^{\nu\rho\tau\omega}+2\mathcal{S}^{\nu\sigma} \mathcal{H}^{\mu\rho\tau\omega}\nonumber\\
&+2\mathcal{S}^{\nu\omega} \mathcal{H}^{\mu\rho\sigma\tau}+2\mathcal{S}^{\mu\omega} \mathcal{H}^{\nu\rho\sigma\tau}+2\mathcal{S}^{\nu\tau} \mathcal{H}^{\mu\rho\sigma\omega}+2\mathcal{S}^{\mu\tau} \mathcal{H}^{\nu\rho\sigma\omega}\nonumber\\
&+2\mathcal{S}^{\sigma\omega} \mathcal{H}^{\mu\nu\rho\tau}+2\mathcal{S}^{\rho\omega} \mathcal{H}^{\mu\nu\sigma\tau}+2\mathcal{S}^{\sigma\tau}\mathcal{H}^{\mu\nu\rho\omega}+2\mathcal{S}^{\rho\tau} \mathcal{H}^{\mu\nu\sigma\omega}\nonumber\\
&-\frac{1}{2}\bigg(\mathcal{H}^{\mu  \rho  \tau}\mathcal{F}^{\nu\sigma\omega}+\mathcal{H}^{\nu  \rho  \tau}\mathcal{F}^{\mu\sigma\omega} +\mathcal{H}^{\mu  \sigma  \tau}\mathcal{F}^{\nu\rho\omega}+\mathcal{H}^{\nu  \sigma  \tau}\mathcal{F}^{\mu\rho\omega}\nonumber\\
&\quad\quad+\mathcal{H}^{\mu  \rho  \omega}\mathcal{F}^{\nu\sigma\tau}+\mathcal{H}^{\nu  \rho  \omega}\mathcal{F}^{\mu\sigma\tau} +\mathcal{H}^{\mu  \sigma  \omega}\mathcal{F}^{\nu\rho\tau}+\mathcal{H}^{\nu  \sigma  \omega}\mathcal{F}^{\mu\rho\tau}\bigg)\nonumber\\
&-\frac{1}{2D}\bigg( \left(\mathcal{F}^{\mu\rho} \mathcal{F}^{\nu\sigma} + \mathcal{F}^{\mu\sigma}\mathcal{F}^{\nu\rho}\right) \mathcal{H}^{\tau\omega}  +\left(\mathcal{F}^{\rho\tau} \mathcal{F}^{\sigma\omega} + \mathcal{F}^{\sigma\tau}\mathcal{F}^{\rho\omega}\right) \mathcal{H}^{\mu\nu}\nonumber\\
&\quad\quad\quad+\left(\mathcal{F}^{\mu\tau} \mathcal{F}^{\nu\omega} + \mathcal{F}^{\nu\tau}\mathcal{F}^{\mu\omega}\right) \mathcal{H}^{\rho\sigma}\bigg)\nonumber\\
&+\frac{3}{32 D^2}\mathcal{F}^\tau\mathcal{F}^\omega\left( \mathcal{F}^{\nu } \mathcal{F}^{\rho } \mathcal{H}^{\mu\sigma}+\mathcal{F}^{\mu } \mathcal{F}^{\sigma }\mathcal{H}^{\nu\rho}+ \mathcal{F}^{\mu } \mathcal{F}^{\rho }  \mathcal{H}^{\nu\sigma}+\mathcal{F}^{\nu } \mathcal{F}^{\sigma }\mathcal{H}^{\mu\rho} \right)\nonumber\\
&+\frac{3}{32 D^2}\mathcal{F}^{\rho } \mathcal{F}^{\sigma }\left(\mathcal{F}^{\mu } \mathcal{F}^{\omega } \mathcal{H}^{\nu  \tau } +\mathcal{F}^{\nu } \mathcal{F}^{\tau } \mathcal{H}^{\mu  \omega }+\mathcal{F}^{\nu }  \mathcal{F}^{\omega }\mathcal{H}^{\mu  \tau }+\mathcal{F}^{\mu } \mathcal{F}^{\tau } \mathcal{H}^{\nu  \omega }\right) \nonumber\\
&+\frac{3}{32 D^2} \mathcal{F}^{\mu } \mathcal{F}^{\nu } \left( \mathcal{F}^{\sigma } \mathcal{F}^{\tau } \mathcal{H}^{\rho \omega } +\mathcal{F}^{\rho } \mathcal{F}^{\omega } \mathcal{H}^{\sigma  \tau }+\mathcal{F}^{\rho } \mathcal{F}^{\tau } \mathcal{H}^{\sigma \omega } +\mathcal{F}^{\sigma} \mathcal{F}^{\omega } \mathcal{H}^{\rho  \tau }  \right)\nonumber\\
& +\frac{1}{8D} p_1^2p_2^2 p_3^2 (p_1^2p_3^2+p_1^2p_2^2+p_2^2 p_3^3)  \eta^{\mu\nu} \eta^{\rho\sigma} \eta^{\tau\omega}\nonumber I^0 \\
&+\frac{1}{32D^2} p_1^4p_3^4 \left(3  \mathcal{F}_\rho \mathcal{F}_\sigma-4Dp_2^\rho p_2^\sigma \right)\eta^{\tau\omega} \eta^{\mu\nu}I^0\nonumber\\
&+\frac{1}{32D^2} p_1^4p_2^4  \left(3 \mathcal{F}_\tau \mathcal{F}_\omega -4D p_3^{\tau} p_3^{\omega}\right)\eta^{\rho\sigma} \eta^{\mu\nu}I^0\nonumber\\
&+\frac{1}{32D^2} p_2^4p_3^4\left(3 \mathcal{F}_\mu \mathcal{F}_\nu -4Dp_1^{\mu} p_1^{\nu}\right)\eta^{\rho\sigma} \eta^{\tau\omega}I^0\nonumber\\
&+\frac{1}{64 D^2}\mathcal{V}^{\mu\nu\rho\sigma\tau\omega} I^0\nonumber
\end{align}
where
\begin{align}\eql{FF}
\mathcal{V}^{\mu\nu\rho\sigma\tau\omega} &=\ 4 \left(\mathcal{F}^{\nu  \rho }\mathcal{F}^{\mu  \sigma }+\mathcal{F}^{\mu  \rho } \mathcal{F}^{\nu  \sigma }\right) \mathcal{F}^\tau \mathcal{F}^\omega\nonumber\\
&+4  \left( \mathcal{F}^{\nu  \tau } \mathcal{F}^{\mu  \omega}+\mathcal{F}^{\mu  \tau } \mathcal{F}^{\nu  \omega}\right) \mathcal{F}^{\rho } \mathcal{F}^{\sigma }\nonumber\\
&+4\left(\mathcal{F}^{\sigma  \tau } \mathcal{F}^{\rho  \omega }+\mathcal{F}^{\rho  \tau } \mathcal{F}^{\sigma \omega }\right)\mathcal{F}^\mu \mathcal{F}^\nu\nonumber\\
&+\frac{3 p_1^4p_2^2p_3^2}{D^2} \bigg(\mathcal{F}^{\sigma }  \mathcal{F}^{\omega }\left(\mathcal{F}^{\nu  \rho } \mathcal{F}^{\mu  \tau }+\mathcal{F}^{\mu  \rho }\mathcal{F}^{\nu  \tau }\right)+(\rho\leftrightarrow\sigma)+ (\tau\leftrightarrow\omega)\bigg)\nonumber\\
&+\frac{3 p_1^2p_2^4p_3^2}{D^2}\bigg(\mathcal{F}^{\nu }\mathcal{F}^{\omega }  \left(\mathcal{F}^{\mu  \sigma } \mathcal{F}^{\rho \tau }+\mathcal{F}^{\mu  \rho } \mathcal{F}^{\sigma  \tau }\right) +(\mu\leftrightarrow\nu)+ (\tau\leftrightarrow\omega)\bigg)\nonumber\\
&+\frac{3 p_1^2p_2^2p_3^4}{D^2}\bigg(\mathcal{F}^{\nu } \mathcal{F}^{\sigma } \left(\mathcal{F}^{\rho  \tau } \mathcal{F}^{\mu  \omega }+\mathcal{F}^{\mu  \tau } \mathcal{F}^{\rho  \omega }\right)+(\mu\leftrightarrow\nu)+(\rho\leftrightarrow\sigma)\bigg)\nonumber\\
&+2p_1^2 p_2^2\left(3 \mathcal{F}^{\tau } \mathcal{F}^{\omega }-4D\mathcal{S}^{\tau\omega}\right)\left( \mathcal{S}^{\nu  \rho }\mathcal{S}^{\mu  \sigma }+ \mathcal{S}^{\mu  \rho }\mathcal{S}^{\nu  \sigma }-\mathcal{S}^{\mu\nu} \mathcal{S}^{\rho\sigma}\right)\nonumber\\
&+2p_1^2 p_3^2\left(3 \mathcal{F}^{\rho } \mathcal{F}^{\sigma }-4D\mathcal{S}^{\rho\sigma}\right)\left(\mathcal{S}^{\nu\tau} \mathcal{S}^{\mu\omega} + \mathcal{S}^{\mu\tau}\mathcal{S}^{\nu\omega}-\mathcal{S}^{\mu\nu} \mathcal{S}^{\tau\omega}\right)\nonumber\\
&+2p_2^2 p_3^3\left(3 \mathcal{F}^{\mu } \mathcal{F}^{\nu }-4D\mathcal{S}^{\mu\nu}\right) \left(\mathcal{S}^{\sigma\tau} \mathcal{S}^{\rho\omega}+\mathcal{S}^{\rho\tau} \mathcal{S}^{\sigma\omega}- \mathcal{S}^{\rho\sigma}\mathcal{S}^{\tau\omega}\right)\nonumber\\
&- \mathcal{F}^{\rho} \mathcal{F}^{\sigma } \mathcal{F}^{\tau } \mathcal{F}^{\omega }\mathcal{S}^{\mu\nu}- \mathcal{F}^{\mu } \mathcal{F}^{\nu } \mathcal{F}^{\rho } \mathcal{F}^{\sigma }\mathcal{S}^{\tau\omega}- \mathcal{F}^{\mu } \mathcal{F}^{\nu } \mathcal{F}^{\tau } \mathcal{F}^{\omega }\mathcal{S}^{\rho\sigma}\nonumber\\
&+\frac{15}{4D} \mathcal{F}^{\mu } \mathcal{F}^{\nu } \mathcal{F}^{\rho } \mathcal{F}^{\sigma } \mathcal{F}^{\tau }\mathcal{F}^{\omega }.
\end{align}
Once again we have fixed the index order and dropped the momentum indices for clarity. Here, as defined in the main text, 
\begin{align}
D &= (p_A\cdot p_B)^2 - p_A^2 p_B^2
\end{align} 
with $A \neq B$ (this definition is independent of the values of $A$ and $B$). The rest of the terms can be generated by taking various combinations of the trace of \Eq{FF}. 
The final result is:

\begin{align}
\aavg{T^{\mu\nu}(p_1) T^{\rho\sigma}&(p_2) T^{\tau\omega}(p_3)}_{\text{vector}}^{(4)}\nonumber\\
 &=\ 8\mathcal{H}^{\mu\nu\rho\sigma\tau\omega}+2\mathcal{S}^{\mu\nu} \mathcal{H}^{\rho\sigma\tau\omega}+2\mathcal{S}^{\rho\sigma}\mathcal{H}^{\mu\nu\tau\omega}+2\mathcal{S}^{\tau\omega}\mathcal{H}^{\mu\nu\rho\sigma}\nonumber\\
&+2\mathcal{S}^{\mu\rho}\mathcal{H}^{\nu\sigma\tau\omega}+2\mathcal{S}^{\nu\rho} \mathcal{H}^{\mu\sigma\tau\omega}+2\mathcal{S}^{\mu\sigma} \mathcal{H}^{\nu\rho\tau\omega}+2\mathcal{S}^{\nu\sigma} \mathcal{H}^{\mu\rho\tau\omega}\nonumber\\
&+2\mathcal{S}^{\nu\omega} \mathcal{H}^{\mu\rho\sigma\tau}+2\mathcal{S}^{\mu\omega} \mathcal{H}^{\nu\rho\sigma\tau}+2\mathcal{S}^{\nu\tau} \mathcal{H}^{\mu\rho\sigma\omega}+2\mathcal{S}^{\mu\tau} \mathcal{H}^{\nu\rho\sigma\omega}\nonumber\\
&+2\mathcal{S}^{\sigma\omega} \mathcal{H}^{\mu\nu\rho\tau}+2\mathcal{S}^{\rho\omega} \mathcal{H}^{\mu\nu\sigma\tau}+2\mathcal{S}^{\sigma\tau}\mathcal{H}^{\mu\nu\rho\omega}+2\mathcal{S}^{\rho\tau} \mathcal{H}^{\mu\nu\sigma\omega}\nonumber\\
&-\frac{1}{2}\bigg(\mathcal{H}^{\mu  \rho  \tau}\mathcal{F}^{\nu\sigma\omega}+\mathcal{H}^{\nu  \rho  \tau}\mathcal{F}^{\mu\sigma\omega} +\mathcal{H}^{\mu  \sigma  \tau}\mathcal{F}^{\nu\rho\omega}+\mathcal{H}^{\nu  \sigma  \tau}\mathcal{F}^{\mu\rho\omega}\nonumber\\
&\ \ \ \ \ +\mathcal{H}^{\mu  \rho  \omega}\mathcal{F}^{\nu\sigma\tau}+\mathcal{H}^{\nu  \rho  \omega}\mathcal{F}^{\mu\sigma\tau} +\mathcal{H}^{\mu  \sigma  \omega}\mathcal{F}^{\nu\rho\tau}+\mathcal{H}^{\nu  \sigma  \omega}\mathcal{F}^{\mu\rho\tau}\bigg)\nonumber\\
&-\frac{1}{2D}\bigg( \left(\mathcal{F}^{\mu\rho} \mathcal{F}^{\nu\sigma} + \mathcal{F}^{\mu\sigma}\mathcal{F}^{\nu\rho}\right) \mathcal{H}^{\tau\omega}  +\left(\mathcal{F}^{\rho\tau} \mathcal{F}^{\sigma\omega} + \mathcal{F}^{\sigma\tau}\mathcal{F}^{\rho\omega}\right) \mathcal{H}^{\mu\nu}\nonumber\\
&\ \ \ \ \  +\left(\mathcal{F}^{\mu\tau} \mathcal{F}^{\nu\omega} + \mathcal{F}^{\nu\tau}\mathcal{F}^{\mu\omega}\right) \mathcal{H}^{\rho\sigma}\bigg)\nonumber\\
&+\frac{3}{32 D^2}\mathcal{F}^\tau\mathcal{F}^\omega\left( \mathcal{F}^{\nu } \mathcal{F}^{\rho } \mathcal{H}^{\mu\sigma}+\mathcal{F}^{\mu } \mathcal{F}^{\sigma }\mathcal{H}^{\nu\rho}+ \mathcal{F}^{\mu } \mathcal{F}^{\rho }  \mathcal{H}^{\nu\sigma}+\mathcal{F}^{\nu } \mathcal{F}^{\sigma }\mathcal{H}^{\mu\rho} \right)\nonumber\\
&+\frac{3}{32 D^2}\mathcal{F}^{\rho } \mathcal{F}^{\sigma }\left(\mathcal{F}^{\mu } \mathcal{F}^{\omega } \mathcal{H}^{\nu  \tau } +\mathcal{F}^{\nu } \mathcal{F}^{\tau } \mathcal{H}^{\mu  \omega }+\mathcal{F}^{\nu }  \mathcal{F}^{\omega }\mathcal{H}^{\mu  \tau }+\mathcal{F}^{\mu } \mathcal{F}^{\tau } \mathcal{H}^{\nu  \omega }\right) \nonumber\\
&+\frac{3}{32 D^2} \mathcal{F}^{\mu } \mathcal{F}^{\nu } \left( \mathcal{F}^{\sigma } \mathcal{F}^{\tau } \mathcal{H}^{\rho \omega } +\mathcal{F}^{\rho } \mathcal{F}^{\omega } \mathcal{H}^{\sigma  \tau }+\mathcal{F}^{\rho } \mathcal{F}^{\tau } \mathcal{H}^{\sigma \omega } +\mathcal{F}^{\sigma} \mathcal{F}^{\omega } \mathcal{H}^{\rho  \tau }  \right)\nonumber\\
&+\frac{1}{64 D^2}\mathcal{V}^{\mu\nu\rho\sigma\tau\omega} I^0.
\end{align}

\newpage
\bibliographystyle{utphys}
\bibliography{mycites}

\providecommand{\href}[2]{#2}\begingroup\raggedright\begin{thebibliography}{10}

\bibitem{Epstein:1965zza}
H.~Epstein, V.~Glaser, and A.~Jaffe, ``{Nonpositivity of energy density in
  Quantized field theories},''
\href{http://dx.doi.org/10.1007/BF02749799}{{\em Nuovo Cim.} {\bfseries 36}
  (1965) 1016}.

\bibitem{Latorre:1997ea}
J.~I. Latorre and H.~Osborn, ``{Modified weak energy condition for the energy
  momentum tensor in quantum field theory.},''
  \href{http://dx.doi.org/10.1016/S0550-3213(97)00667-6}{{\em Nucl. Phys.}
  {\bfseries B511} (1998) 737--759},
\href{http://arxiv.org/abs/hep-th/9703196}{{\ttfamily arXiv:hep-th/9703196
  [hep-th]}}.

\bibitem{Ford:1990id}
L.~H. Ford, ``{Constraints on negative energy fluxes},''
\href{http://dx.doi.org/10.1103/PhysRevD.43.3972}{{\em Phys. Rev.} {\bfseries
  D43} (1991) 3972--3978}.

\bibitem{Ford:1994bj}
L.~H. Ford and T.~A. Roman, ``{Averaged energy conditions and quantum
  inequalities},'' \href{http://dx.doi.org/10.1103/PhysRevD.51.4277}{{\em Phys.
  Rev.} {\bfseries D51} (1995) 4277--4286},
\href{http://arxiv.org/abs/gr-qc/9410043}{{\ttfamily arXiv:gr-qc/9410043
  [gr-qc]}}.

\bibitem{Ford:1996er}
L.~H. Ford and T.~A. Roman, ``{Restrictions on negative energy density in flat
  space-time},'' \href{http://dx.doi.org/10.1103/PhysRevD.55.2082}{{\em Phys.
  Rev.} {\bfseries D55} (1997) 2082--2089},
\href{http://arxiv.org/abs/gr-qc/9607003}{{\ttfamily arXiv:gr-qc/9607003
  [gr-qc]}}.

\bibitem{Ford:1978qya}
L.~H. Ford, ``{Quantum Coherence Effects and the Second Law of
  Thermodynamics},''
\href{http://dx.doi.org/10.1098/rspa.1978.0197}{{\em Proc. Roy. Soc. Lond.}
  {\bfseries A364} (1978) 227--236}.

\bibitem{Morris:1988tu}
M.~S. Morris, K.~S. Thorne, and U.~Yurtsever, ``{Wormholes, Time Machines, and
  the Weak Energy Condition},''
\href{http://dx.doi.org/10.1103/PhysRevLett.61.1446}{{\em Phys. Rev. Lett.}
  {\bfseries 61} (1988) 1446--1449}.

\bibitem{Curiel:2014zba}
E.~Curiel, ``{A Primer on Energy Conditions},''
\href{http://arxiv.org/abs/1405.0403}{{\ttfamily arXiv:1405.0403
  [physics.hist-ph]}}.

\bibitem{Osborn:1993cr}
H.~Osborn and A.~C. Petkou, ``{Implications of conformal invariance in field
  theories for general dimensions},''
  \href{http://dx.doi.org/10.1006/aphy.1994.1045}{{\em Annals Phys.} {\bfseries
  231} (1994) 311--362},
\href{http://arxiv.org/abs/hep-th/9307010}{{\ttfamily hep-th/9307010}}.

\bibitem{Blanco:2013lea}
D.~D. Blanco and H.~Casini, ``{Localization of Negative Energy and the
  Bekenstein Bound},''
  \href{http://dx.doi.org/10.1103/PhysRevLett.111.221601}{{\em Phys. Rev.
  Lett.} {\bfseries 111} no.~22, (2013) 221601},
\href{http://arxiv.org/abs/1309.1121}{{\ttfamily arXiv:1309.1121 [hep-th]}}.

\bibitem{Hofman:2008ar}
D.~M. Hofman and J.~Maldacena, ``{Conformal collider physics: Energy and charge
  correlations},'' \href{http://dx.doi.org/10.1088/1126-6708/2008/05/012}{{\em
  JHEP} {\bfseries 05} (2008) 012},
\href{http://arxiv.org/abs/0803.1467}{{\ttfamily arXiv:0803.1467}}.

\bibitem{Zhiboedov:2013opa}
A.~Zhiboedov, ``{On Conformal Field Theories With Extremal $a/c$ Values},''
  \href{http://dx.doi.org/10.1007/JHEP04(2014)038}{{\em JHEP} {\bfseries 04}
  (2014) 038},
\href{http://arxiv.org/abs/1304.6075}{{\ttfamily arXiv:1304.6075 [hep-th]}}.

\bibitem{Abreu:1990us}
{\bfseries DELPHI} Collaboration, P.~Abreu {\em et~al.}, ``{Energy-energy
  correlations in hadronic final states from Z0 decays},''
\href{http://dx.doi.org/10.1016/0370-2693(90)91097-U}{{\em Phys. Lett.}
  {\bfseries B252} (1990) 149--158}.

\bibitem{Akrawy:1990hy}
{\bfseries OPAL} Collaboration, M.~Z. Akrawy {\em et~al.}, ``{A Measurement of
  energy correlations and a determination of alpha-s (M2 (Z0)) in e+ e-
  annihilations at s**(1/2) = 91-GeV},''
\href{http://dx.doi.org/10.1016/0370-2693(90)91098-V}{{\em Phys. Lett.}
  {\bfseries B252} (1990) 159--169}.

\bibitem{Abe:1994wv}
{\bfseries SLD} Collaboration, K.~Abe {\em et~al.}, ``{Measurement of alpha-s
  from energy-energy correlations at the Z0 resonance},''
  \href{http://dx.doi.org/10.1103/PhysRevD.50.5580}{{\em Phys. Rev.} {\bfseries
  D50} (1994) 5580--5590},
\href{http://arxiv.org/abs/hep-ex/9405006}{{\ttfamily arXiv:hep-ex/9405006
  [hep-ex]}}.

\bibitem{Penrose:1993ud}
R.~Penrose, R.~D. Sorkin, and E.~Woolgar, ``{A Positive mass theorem based on
  the focusing and retardation of null geodesics},''
\href{http://arxiv.org/abs/gr-qc/9301015}{{\ttfamily arXiv:gr-qc/9301015
  [gr-qc]}}.

\bibitem{Gao:2000ga}
S.~Gao and R.~M. Wald, ``{Theorems on gravitational time delay and related
  issues},'' \href{http://dx.doi.org/10.1088/0264-9381/17/24/305}{{\em Class.
  Quant. Grav.} {\bfseries 17} (2000) 4999--5008},
\href{http://arxiv.org/abs/gr-qc/0007021}{{\ttfamily arXiv:gr-qc/0007021
  [gr-qc]}}.

\bibitem{Klinkhammer:1991ki}
G.~Klinkhammer, ``{Averaged energy conditions for free scalar fields in flat
  space-times},''
\href{http://dx.doi.org/10.1103/PhysRevD.43.2542}{{\em Phys. Rev.} {\bfseries
  D43} (1991) 2542--2548}.

\bibitem{Wald:1991xn}
R.~M. Wald and U.~Yurtsever, ``{General proof of the averaged null energy
  condition for a massless scalar field in two-dimensional curved
  space-time},''
\href{http://dx.doi.org/10.1103/PhysRevD.44.403}{{\em Phys. Rev.} {\bfseries
  D44} (1991) 403--416}.

\bibitem{Kelly:2014mra}
W.~R. Kelly and A.~C. Wall, ``{Holographic proof of the averaged null energy
  condition},'' \href{http://dx.doi.org/10.1103/PhysRevD.90.106003,
  10.1103/PhysRevD.91.069902}{{\em Phys. Rev.} {\bfseries D90} no.~10, (2014)
  106003}, \href{http://arxiv.org/abs/1408.3566}{{\ttfamily arXiv:1408.3566
  [gr-qc]}}.
[Erratum: Phys. Rev.D91,no.6,069902(2015)].

\bibitem{Hofman:2009ug}
D.~M. Hofman, ``{Higher Derivative Gravity, Causality and Positivity of Energy
  in a UV complete QFT},''
  \href{http://dx.doi.org/10.1016/j.nuclphysb.2009.08.001}{{\em Nucl. Phys.}
  {\bfseries B823} (2009) 174--194},
\href{http://arxiv.org/abs/0907.1625}{{\ttfamily arXiv:0907.1625 [hep-th]}}.

\bibitem{Luscher:1974ez}
M.~Luscher and G.~Mack, ``{Global Conformal Invariance in Quantum Field
  Theory},''
\href{http://dx.doi.org/10.1007/BF01608988}{{\em Commun. Math. Phys.}
  {\bfseries 41} (1975) 203--234}.

\end{thebibliography}\endgroup

\end{document}